\documentclass[fleqn,usenatbib]{mnras}

\usepackage{newtxtext,newtxmath}

\usepackage[T1]{fontenc}
\usepackage{ae,aecompl}
 
\usepackage{natbib}
\bibpunct{(}{)}{;}{a}{}{,}

\usepackage{amsmath}
\usepackage{amsfonts}
\usepackage{amssymb}
\usepackage{wasysym}
\usepackage{textcomp}

\usepackage{booktabs}
\usepackage{graphicx}
\usepackage{color}

\newcommand{\be}{\begin{equation}}
\newcommand{\ee}{\end{equation}}
\newcommand{\bea}{\begin{eqnarray}}
\newcommand{\eea}{\end{eqnarray}}

\newcommand{\m}{\,\hbox{m}}
\newcommand{\mm}{\,\hbox{mm}}
\newcommand{\km}{\,\hbox{km}}
\newcommand{\mum}{\,\hbox{\textmu{}m}}
\newcommand{\cm}{\,\hbox{cm}}

\newcommand{\AU}{\,\hbox{AU}}
\newcommand{\g}{\,\hbox{g}}
\newcommand{\s}{\,\hbox{s}}

\newcommand{\Myr}{\,\hbox{Myr}}
\newcommand{\Gyr}{\,\hbox{Gyr}}

\newcommand{\erg}{\,\hbox{erg}}
\newcommand{\K}{\,\hbox{K}}

\title[Debris Disc Constraints on Planetesimals]{Debris Disc Constraints on Planetesimal Formation}

\author[A. V. Krivov et al.]{Alexander V. Krivov,$^{1}$\thanks{E-mail: krivov@astro.uni-jena.de (AVK)}
Aljoscha Ide,$^{1}$
Torsten L\"ohne,$^{1}$
Anders Johansen,$^{2}$
\newauthor
and J\"urgen Blum$^{3}$
\\
$^{1}$Astrophysikalisches Institut und Universit\"atssternwarte, Friedrich-Schiller-Universit\"at Jena,
      Schillerg\"a{\ss}chen~2--3, 07745 Jena, Germany\\
$^{2}$Lund Observatory, Department of Astronomy and Theoretical Physics,
      Lund University, Box 43, SE-221 00 Lund, Sweden\\
$^{3}$Institut f\"ur Geophysik und extraterrestrische Physik,
      Technische Universtit\"at Braunschweig, Mendelssohnstr. 3,
      38106, Braunschweig,\\ Germany
}

\date{Accepted 2017 November 8. Received 2017 November 8; in original form 2017 September 5}

\pubyear{2017}

\begin{document}
\label{firstpage}
\pagerange{\pageref{firstpage}--\pageref{lastpage}}
\maketitle

\begin{abstract}
Two basic routes for planetesimal formation have been proposed over the last few
decades.  One is a classical ``slow-growth'' scenario.  Another one is
particle concentration models, in which small pebbles are concentrated
locally and then collapse gravitationally to form planetesimals.  Both types
of models make certain predictions for the size spectrum and internal structure
of newly-born planetesimals.  We use these predictions
as input to simulate collisional evolution of debris discs left
after the gas dispersal.  The debris disc emission as a function of a
system's age computed in these simulations is compared with several Spitzer
and Herschel debris disc surveys around A-type stars.  We confirm that the
observed brightness evolution for the majority of discs can be reproduced by
classical models.  Further, we find that it is equally consistent with
the size distribution of planetesimals predicted by particle concentration models~---
provided the objects are loosely bound ``pebble piles'' as these models also predict.
Regardless of the assumed planetesimal formation mechanism, explaining the brightest
debris discs in the samples uncovers a ``disc mass problem.''
To reproduce such discs by collisional simulations,
a total mass of planetesimals of up to $\sim 1000$ Earth masses is required,
which exceeds the total mass of solids available in the protoplanetary
progenitors of debris discs.  This may indicate that stirring was delayed in
some of the bright discs, that giant impacts occurred recently in some of
them, that some systems may be younger than previously thought, or
that non-collisional processes contribute significantly to the dust
production.
\end{abstract}



 
\section{Introduction}

Debris discs are belts of leftover planetesimals, i.e., comets and asteroids, around stars
\citep{wyatt-2008,krivov-2010,matthews-et-al-2013}.
They are commonly observed through the thermal emission of the dust that these small bodies produce in
collisions and other destructive processes.
To get insights into the properties of directly invisible planetesimals, observed emission is interpreted
by collisional modelling of planetesimal belts, supplemented by thermal emission calculations
\citep[e.g.][]{krivov-et-al-2008}.
A number of prominent discs have been modelled individually this way
\citep[e.g.][]{wyatt-et-al-1999,wyatt-dent-2002,thebault-et-al-2003,mueller-et-al-2009,%
reidemeister-et-al-2011,loehne-et-al-2011,%
schueppler-et-al-2014,schueppler-et-al-2015,schueppler-et-al-2016}.
Larger samples of discs have been modelled on a statistical basis
\citep[e.g.][]{wyatt-et-al-2007b,kennedy-wyatt-2010,gaspar-et-al-2012,gaspar-et-al-2013,pawellek-et-al-2014,pawellek-krivov-2015,geiler-krivov-2017}.
All these models are able to reproduce the available data and the observed statistical trends 
reasonably well.

These models, however, typically assume all of the solids to initially 
have a single power-law size distribution $dN/dm \propto m^{-\alpha}$ 
with a slope close to $\alpha \approx 1.8$...$1.9$. 
A power law with such a slope, which we call ``the Dohnanyi slope'', 
is an analytic solution for the steady-state size distribution found
for an idealized collisional cascade
\citep{dohnanyi-1969,o'brien-greenberg-2003,wyatt-et-al-2011}.
The results of elaborate numerical simulations with more realistic physics are expected to 
deviate from this power law only slightly.
This explains why taking it as an initial condition for numerical simulations is so 
popular: it ensures fast convergence towards the exact distribution.
However, the convenient assumption of an initial Dohnanyi-like slope ignores the fact that
the initial size distribution of freshly-formed planetesimals is likely to be very different
from Dohnanyi's. 
As time elapses, larger and larger planetesimals get involved in the collisional cascade.
Thus taking another initial distribution of them should affect the modelling results
\citep[][]{loehne-et-al-2007}.

That primordial size distribution remains poorly known, because so is the process by which planetesimals form.
Two basic scenarios have been proposed \citep[see, e.g.,][for a recent review]{johansen-et-al-2014}.
One is a classical scenario of a slow incremental growth, first by 
fluffy agglomeration \citep[e.g.,][]{wada-et-al-2009,okuzumi-et-al-2012,wada-et-al-2013}
or mass transfer from small to large aggregates \citep[e.g.,][]{wurm-et-al-2005,windmark-et-al-2012} and
then by gravity-driven collisional assemblage of larger planetesimals
\citep[e.g.,][]{kenyon-luu-1999b,kenyon-bromley-2008,kobayashi-et-al-2010b,%
kobayashi-et-al-2016}.
Another one is the particle concentration models in which small pebbles are concentrated
locally in eddies, pressure bumps or vortices of a turbulent disc 
\citep[e.g.][]{haghighipour-boss-2003,cuzzi-et-al-2008,johansen-et-al-2009,cuzzi-et-al-2010}
or by streaming instability
\citep[e.g.][]{johansen-et-al-2007,johansen-et-al-2015,simon-et-al-2016,simon-et-al-2017}
and then collapse gravitationally to form planetesimals.
Both types of models
make specific predictions for the size distribution of forming planetesimals,
suggesting it to be shallower than Dohnanyi's.
They also have implications for the internal structure of planetesimals.
For instance, the particle concentration models
by \citet{johansen-et-al-2015} and \citet{simon-et-al-2016,simon-et-al-2017} independently 
propose $\alpha \approx 1.6$  over the size range from a few kilometers to a 
few hundred kilometers and predict planetesimals to be low-density, 
porous bodies.

The goal of this paper is to check what would happen if we took these 
predictions by the planetesimal formation models and used them as initial 
conditions for collisional modelling of debris discs. Would the results still be consistent with debris 
disc observational data?
By comparing the models with the data, we will try to further constrain the initial
size distribution and internal structure of planetesimals. 
Our intention in this work is somewhat similar to the study of \citet{kenyon-bromley-2008} and
\citet{kobayashi-loehne-2014} who investigated the debris disc evolution, assuming classical ``slow growth'' 
planetesimal formation models. For
the Solar System, work along these lines has been done by \citet{morbidelli-et-al-2009}
who showed the size distribution in the asteroid belt to be consistent with ``asteroids born big'',
hence favouring the particle concentration mechanism of asteroid formation.
\citet{vitense-et-al-2012} came to the same conclusion for the Kuiper belt.

Section~2 presents the collisional model and
Section~3 describes the observational datasets on debris discs.
Section~4 compares the model predictions to the observational data.
Section~5 discusses the results.
Section~6 draws our conclusions.

\section{Model}

\subsection{The code}

To model the collisional evolution of the discs,
we use the collisional code ACE \citep[``Analysis of Collisional Evolution'';][]
{krivov-et-al-2006,loehne-et-al-2007,loehne-et-al-2011,krivov-et-al-2013,loehne-et-al-2017}.
The code assumes a certain central star and a planetesimal disc of certain size, mass, excitation and
other properties as input.
The code simulates the collisional evolution of the debris disc,
predicting the coupled radial and size distributions of solids~--- from planetesimals to dust~---
at different time instants.
To compare the simulation results to the observational data, the ACE output is then used to calculate thermal 
emission of dust. Specifically, we compute the thermal emission fluxes at selected far-infrared wavelengths
as functions of system's age.

\subsection{General parameters}

Since resolved images of the vast majority of debris discs show them to be relatively narrow rings
with a typical distance from the star of $\sim 100\AU$ \citep[e.g.][]{pawellek-et-al-2014},
the reference disc model in all ACE runs was a $10\AU$ wide ring around a distance of $100\AU$.
As a central star, we chose an A-type star of $2.16$ solar masses with a luminosity of
$27.7$ times the solar luminosity.
The above choices are close to the median values of the samples described below.
We are aware, of course, that all these parameters for individual discs in those samples vary considerably.
However, we decided to fix them to avoid dealing with an unmanageable set of free parameters and to keep our treatment
as simple as possible.
This was backed up by several tests to see how the debris disc fluxes and their evolution in time
would alter if we varied the luminosity and mass of the host stars, as well as the radius and
width of the debris rings.
The changes were found to be
moderate. The largest quantitative changes arise from variation of the disc radius, consistently with
previous studies \citep[see, e.g., top right panel of Fig. 11 in][]{loehne-et-al-2007}.

We postulated a disc of planetesimals with eccentricities between 0 and 0.1 and inclinations from 0 to 3 
degrees.
Here too, we argue that variation of the assumed eccentricity and inclination distribution would not alter 
the results markedly.
Indeed, the collision rates are nearly independent of eccentricities. This is 
because collisional rates are proportional to the ratio collisional velocity / collisional interaction 
volume. Both quantities are proportional to eccentricity, which then cancels out \citep[see, e.g.,][]{krivov-et-al-2006}.
The same applies to inclinations.
What the eccentricities and inclinations do affect is the collisional velocities.
As a result, the dust production rate at higher eccentricities is higher early on, and it
decays more rapidly at later times.
Quantitatively, however, the effect is only moderate. See, for instance,
Fig.~11 (bottom right) in \citet{loehne-et-al-2007}, which shows the evolution for $e=0.05$, $0.10$, 
$0.15$, and $0.20$, and a discussion therein.

The solids were assumed to consist of an astrosilicate--water ice mixture in a 30\%--70\% ratio.
Since the material composition of extrasolar comets is not known, this choice is arbitrary.
We only made it because silicate--ice mixtures worked better than pure silicate or pure ice
in SED fitting of several debris discs studied previously
\citep[e.g.,][]{reidemeister-et-al-2011,lebreton-et-al-2012,donaldson-et-al-2013,schueppler-et-al-2016}.
Alternatively, we could have used other compositions,
such as mixtures of silicates, ices, carbonaceous materials, and vacuum 
\citep[e.g.,][]{lebreton-et-al-2012,donaldson-et-al-2013} or those
involving hydrocarbons that were found to be the major constituent of the comet 67/P in the Solar System
\citep{fulle-et-al-2016}. On any account, we do not expect the results to be very sensitive to the material
composition chosen.
To avoid unnecessary complications, the Poynting-Robertson drag and stellar wind drag were switched off.
The solids spanned the size range from $s_\text{min}=0.2\mum$ to $s_\text{max}=200\km$.

\subsection{Critical fragmentation energy}

\begin{figure}
\centering
\includegraphics[width=0.49\textwidth, angle=0]{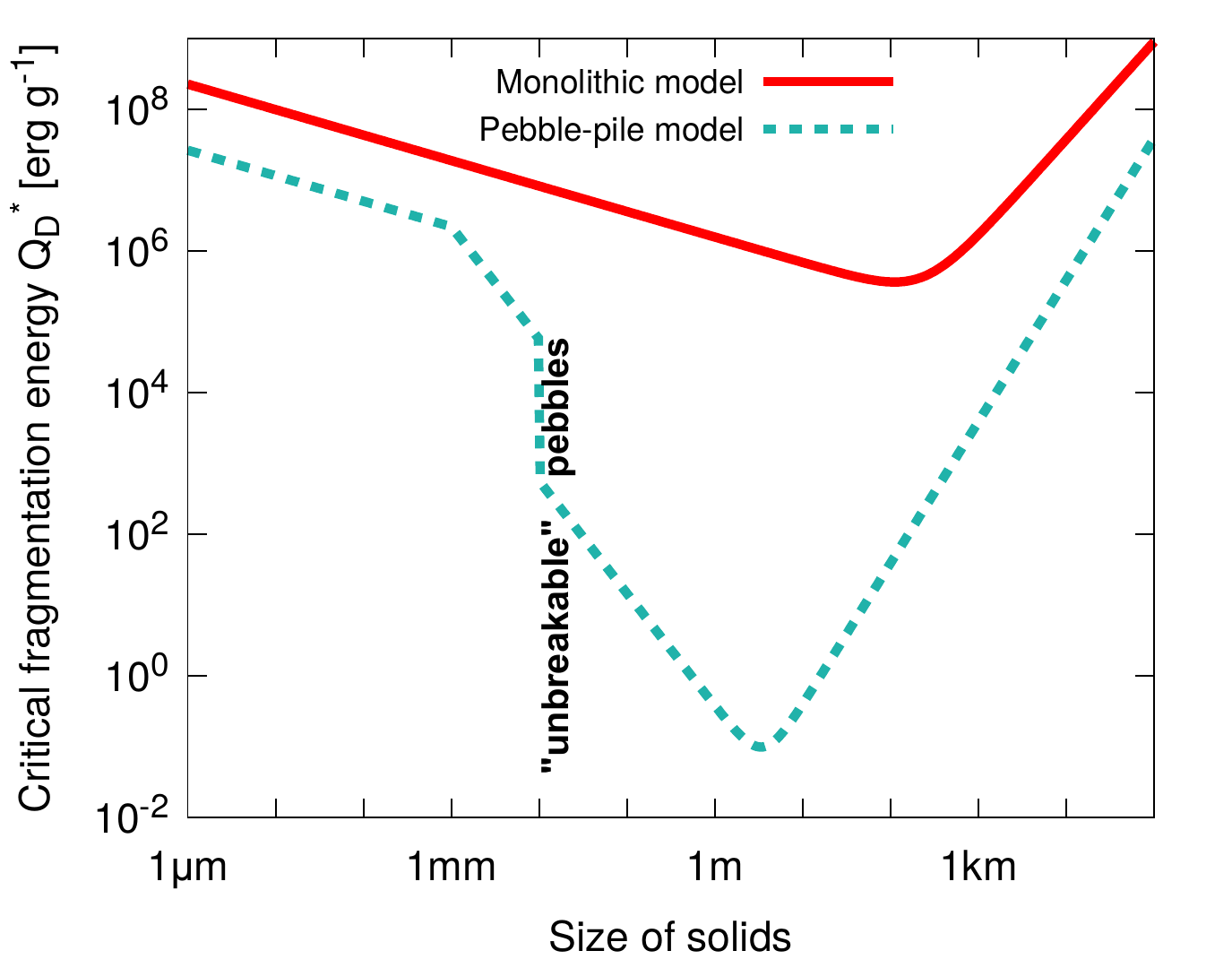}
\caption{
Critical fragmentation energy as a function of size
(for $v_{\text{imp}}=300\m\s^{-1}$).
Red solid line: standard ``monolithic'' model, Eq.~(\ref{eq:QD}).
Cyan dashed line: pebble-pile model, Eq.~(\ref{eq:QD_pp}).
\label{fig:QD}
   }
\end{figure}

To model the outcomes of collisions between the objects, we also needed to assume
a certain critical fragmentation energy, which depends on the material strength,
the object size, and the impact velocity $v_{\text{imp}}$.
We used two prescriptions. 
One of them is appropriate for ``monolithic'', collisionally strong planetesimals 
in the ``slow growth'' planetesimal formation models and is taken
from \citet{schueppler-et-al-2016}:
\be
Q_\text{D}^{\ast}=
\left[
A_{\text{s}}\left(s \over 1\m \right)^{3b_{\text{s}}}
+
A_{\text{g}}\left(s \over 1\km\right)^{3b_{\text{g}}}
\right]
\left(
v_{\text{imp}} \over v_0
\right)^{0.5},
\label{eq:QD}
\ee
where we set $v_0 = 3\km\s^{-1}$ and chose $b_{\text{s}}=-0.12$, $b_{\text{g}}=0.46$,
and $A_{\text{g}}=A_{\text{s}}=5\times10^6 \erg\g^{-1}$.
This prescription is essentially based on the SPH simulations by \citet{benz-asphaug-1999} 
for basalt, complemented with the velocity dependence taken from \citet{stewart-leinhardt-2009}.
The resulting critical fragmentation energy
is plotted in Fig.~\ref{fig:QD}.

In addition to prescription~(\ref{eq:QD}),
we also tried an alternative model for 
$Q_\text{D}^{\ast}$ that we deem more suitable for planetesimals forming through
the particle concentration mechanism.
In this scenario,
planetesimals are likely piles of cm-sized pebbles \citep{johansen-et-al-2015}.
Based on recent laboratory experiments on the strength of dust aggregates
\citep{bukhari-et-al-2017,whizin-et-al-2017},
we can formulate the following prescription for $Q_\text{D}^\ast$:
\be
Q_\text{D}^{\ast}
  = A_\text{pp}\left(\frac{s}{\text{1~mm}}\right)^{3b_\text{pp}}
    \left(
      \frac{v_\text{imp}}{v_0}
    \right)^{0.5}
  + \frac{3Gm}{5s},
\label{eq:QD_pp}
\ee
where
\be
  A_\text{pp} = \left\{\begin{array}{l}
    7\times 10^6~\text{erg/g}~\text{for}~s < 1~\text{cm},\\
    7\times 10^4~\text{erg/g}~\text{for}~s > 1~\text{cm}
                     \end{array}\right.
\label{eq:A_pp}
\ee
and
\be
  b_\text{pp} = \left\{\begin{array}{l}
    -0.12~\text{for}~s < 1~\text{mm},\\
    -0.53~\text{for}~s > 1~\text{mm}.
                     \end{array}\right.
\label{eq:b_pp}
\ee
In the last term in Eq.~(\ref{eq:QD_pp}),
$G$ is the gravitational constant, and $m$ is the mass of an object of radius $s$.
We arrived at Eq.~(2) by using the velocity scaling derived by 
\citet{bukhari-et-al-2017}, i.e.,
$Q_\text{D}^\ast \propto s^{3b_\text{pp}} \propto s^{-1.58}$, 
measured for pebble piles that were compacted to rather homogeneous dust aggregates
with filling factors of 0.35.
To account for the much lower cohesion within uncompacted (i.e., primordial) pebble piles, we 
scaled the pre-factor $A_\text{pp}$ in Eq.~(\ref{eq:QD_pp})
accordingly (see Eq.~(\ref{eq:A_pp}) and Fig.~\ref{fig:QD}).
Our scaling is based on the experimental data by \citet{whizin-et-al-2017} and \citet{bukhari-et-al-2017}.
\citet{whizin-et-al-2017} experimentally investigated the collision 
behaviour of cm-sized pebble clusters.
Extrapolating their data to collisions among equal-sized clusters yields a 
fragmentation threshold of $\sim 1~\cm\s^{-1}$.
This is well below the respective value for compacted dust aggregates, 
which was determined by \citet{bukhari-et-al-2017} to be on the order of $1\m\s^{-1}$. 

When using the model (\ref{eq:QD_pp}),
the size distribution of fragments resulting from the disruption of pebble aggregates was
assumed to have a lower cutoff at the pebble size. To describe the disruption of pebbles
themselves, we used the fragment size distribution 
extending down to sub-micron-sized dust.

In Fig.~\ref{fig:QD}, we compare this
pebble-pile $Q_\text{D}^*(s)$ model with the monolithic one.
The critical fragmentation energy given by
Eq.~(\ref{eq:QD_pp}) falls off more steeply
for $s > 1$~mm (with $b_\text{pp} = -0.53$) and features an additional
drop by two orders of magnitude at the boundary between pebbles ($s
\lesssim 1$~cm) and pebble piles ($s \gtrsim 1$~cm).
Altogether, it is seen that the pebble-pile model predicts much weaker planetesimals with radii between
the pebble size and about one kilometer.
The change in the material strength at the dust sizes is less than by 
order of magnitude.
This, rather moderate,  change reflects a difference between the simulation 
results by 
\citet{benz-asphaug-1999} and laboratory measurements by \citet{whizin-et-al-2017}, who also assumed 
different materials and compositions.
Finally, in the gravity regime, the strength of pebble-pile planetesimals differs
from that of monolithic objects, too \citep{stewart-leinhardt-2009,leinhardt-stewart-2012}.
This is because the propagation of shock waves triggered by the impact through the target and energy 
dissipation within the target in these two cases are different.
As a result, the critical fragmentation energy of pebble-pile objects is dominated by pure
gravitational binding energy (the last term in Eq.~(\ref{eq:QD_pp})).
This leads to $Q_\text{D}^\ast \propto s^2$ for $s \gtrsim 10$~m.

\subsection{Initial size distribution}

\begin{figure}
\centering
\includegraphics[width=0.49\textwidth, angle=0]{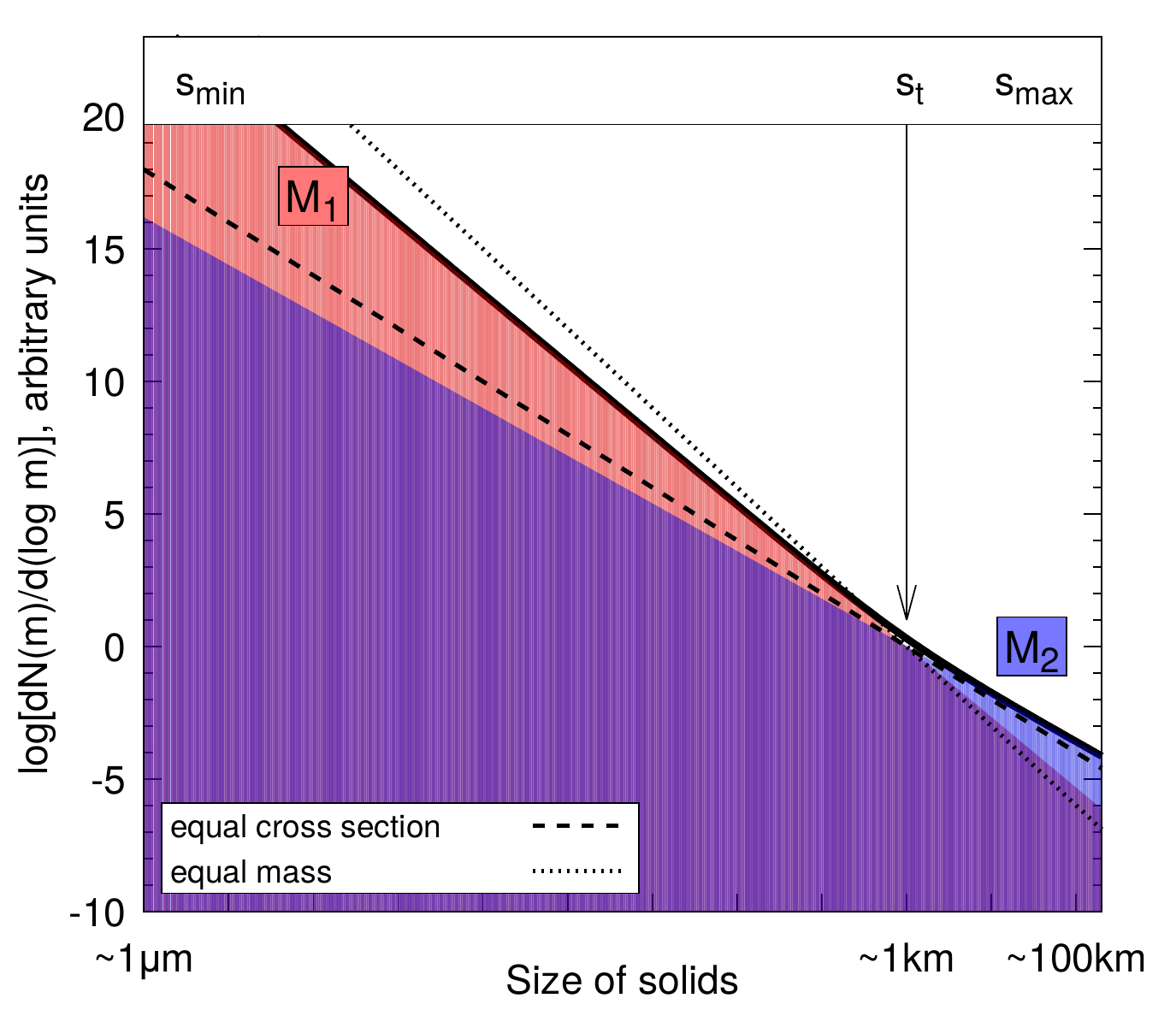}\\
\caption{
Initial size distribution of planetesimals assumed in the collisional simulations (black solid line).
It involves two slopes: a steeper one for small solids of sizes $s_\text{min}$ to $s_\text{t}$ (population~1,
red-filled area) and a flatter one
appropriate for large objects with sizes from $s_\text{t}$ to $s_\text{max}$ (population~2, blue-filled area).
Dashed line indicates a slope ($\alpha=5/3$), for which the {\em cross section} of solids would be the same
in all logarithmic size (or mass) bins. Similarly, dotted line marks the slope ($\alpha=2$), for which
all logarithmic size bins would carry equal {\em mass} of material.
\label{fig:schematic}
   }
\end{figure}

Given the goals of this study, 
the initial size distribution was modelled with
two power laws, a steeper one below a certain size and a shallower one above it
(Fig.~\ref{fig:schematic}).
The first one was assumed to have a Dohnanyi-like slope
$\alpha = (11+3b_\text{s})/(6+3b_\text{s}) = 1.88$ \citep{o'brien-greenberg-2003}.
Here, $b_\text{s}$ is the slope of $Q_D^\star (s)$ in the strength regime (see Eq.~\ref{eq:QD}).
The second power law was a proxy for the primordial size distribution of large planetesimals
right after the gas dispersal.
Accordingly, we used $\alpha=1.6$ as expected in both planetesimal formation scenarios considered 
here.

Since the ACE code does not allow for using broken power-laws, the technical implementation 
was slightly different. We took two initial populations of solids at the same spatial location~--- 
one with a steeper and another one with a shallower power law, as described above. 
Of course, both populations were let to interact with each other collisionally.

\begin{figure}
\centering
\includegraphics[width=0.49\textwidth, angle=0]{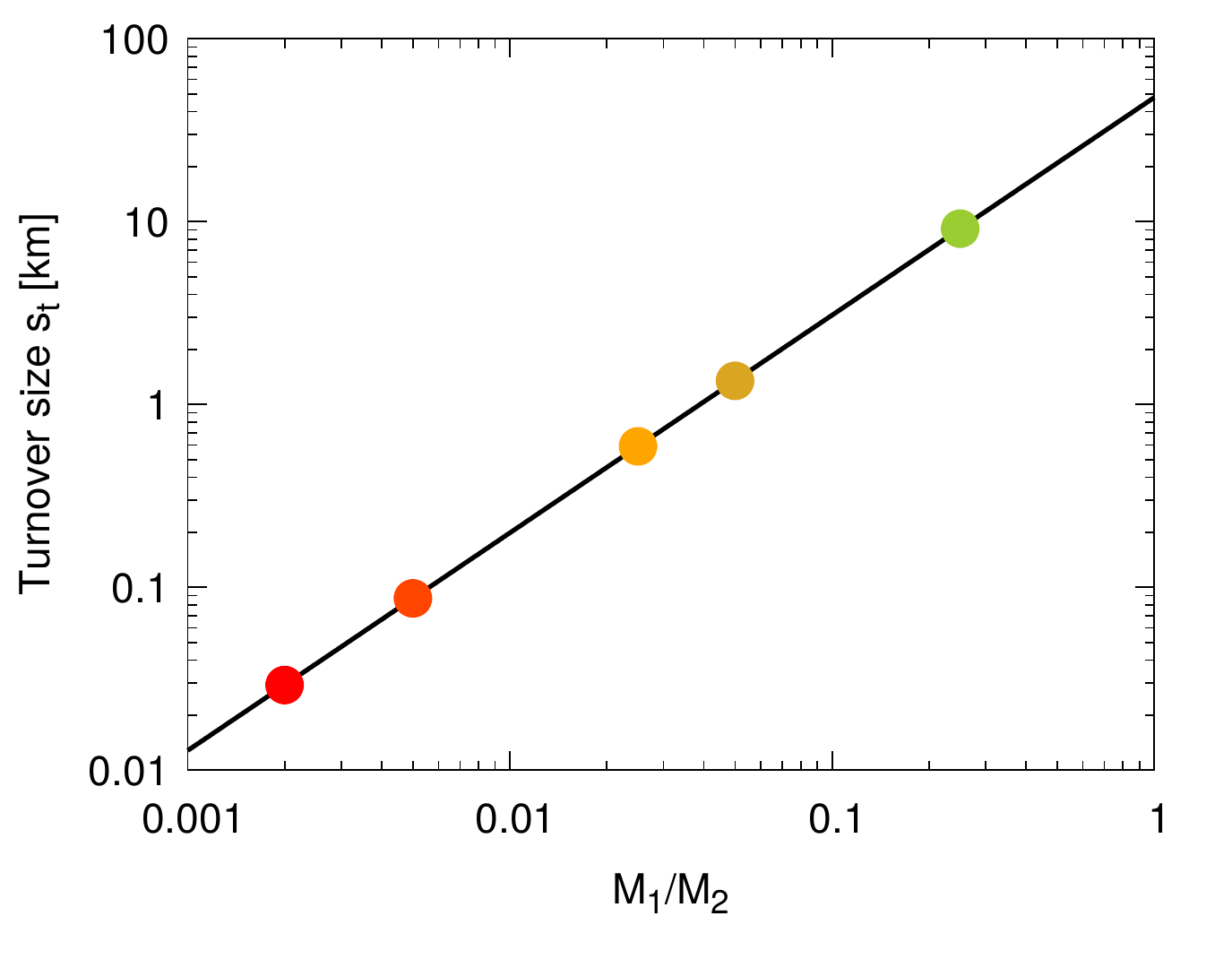}
\caption{
Planetesimals' turnover size as a function of $M_1/M_2$.
Bold dots mark mass ratios $M_1/M_2 = 1/4$, $1/20$, $1/40$, $1/200$, and $1/500$ assumed to produce the
``evolutionary tracks'' in Fig.~\ref{fig:constant mass}.
Each mass ratio is associated with a certain colour; the same colour coding is consistently used throughout the paper.
\label{fig:turnover}
   }
\end{figure}

This two-population setup has two free parameters: the total initial disc mass, $M_1+M_2$,
and the mass ratio of the two populations, $M_1/M_2$.
Fixing a  ratio $M_1/M_2$ is equivalent to selecting
a certain ``turnover'' size $s_\text{t}$, at which the two populations contain
equal numbers of planetesimals (Fig.~\ref{fig:turnover}).
Varying both parameters, and selecting either
Eq.~(\ref{eq:QD}) or Eq.~(\ref{eq:QD_pp}) for $Q_\text{D}^*(s)$,
we try to reproduce the observed level of disc brightness
in the far-infrared and its decay towards older ages.

It is easy to see that different combinations of $M_1/M_2$ (or $s_\text{t}$) together with
different $Q_\text{D}^*(s)$ prescriptions allow us to test both ``slow-growth''
and ``particle concentration'' models of planetesimal formation:
\begin{enumerate}
\item[]
(1) In the slow-growth scenario, growth of larger bodies in a protoplanetary disc is always accompanied 
with fragmentation at smaller sizes. As the gas density goes down, the collisional cascade induced by growing 
embedded stirrers produces ever smaller solids, so that by the time when the disc 
gets gas-free all solids down to dust sizes are generated. As a result, we expect a Dohnanyi-like slope
at sizes below about a kilometer and a shallower slope for larger bodies that have grown in the disc.
Those bodies are expected to be relatively compact (``monolithic''). This means 
that our two-population setup with $s_\text{t}$ in the kilometer range, together with
Eq.~(\ref{eq:QD}) for $Q_\text{D}^*(s)$, is a proxy for the ``slow growth'' scenario.
\item[]
(2) In the particle concentration scenario, the expectation is different. Pebbles are efficiently consumed 
to quickly build large planetesimals with sizes above a few kilometers with a shallower size distribution. 
These will be loosely bound, having a pebble-pile structure. Fragmentation only plays a minor 
role, and by the moment of gas dispersal no Dohnanyi-like tail of subkilometer-sized solids is predicted.
It terms of our two-population setup, population 1 will be nearly absent (very small $M_1/M_2$, or 
$s_\text{t} \ll 1\km$). Thus taking a small $s_\text{t}$ together with
Eq.~(\ref{eq:QD_pp}) for $Q_\text{D}^*(s)$ serves as a proxy for the ``particle concentration'' scenario.
\item[]
(3) With the same setup, we can also test the traditional debris disc models that do not utilise predictions 
of planetesimal formation models at all, just using a Dohnanyi-like initial size distribution across the entire 
size range. 
To this end, it is sufficient to discard population~2 (by setting $M_1/M_2$ to infinity, or $s_\text{t}$ to
$s_\text{max}$) and use Eq.~(\ref{eq:QD}) for $Q_\text{D}^*(s)$.
\end{enumerate}

\subsection{Typical simulation results}

We have performed several ACE simulations with fiducial discs.
In all these runs, the disc mass $M_1+M_2$ was arbitrarily set to the same value
of $100 M_\oplus$, while the mass ratio between the two populations $M_1/M_2$ was varied from
infinity (no population~2, ``traditional'' choice) to 1/500 (almost all the mass in population~2).
All of the runs were done with monolithic planetesimals, except for one where
we assumed pebble piles.

\begin{figure}
\centering
\includegraphics[width=0.49\textwidth, angle=0]{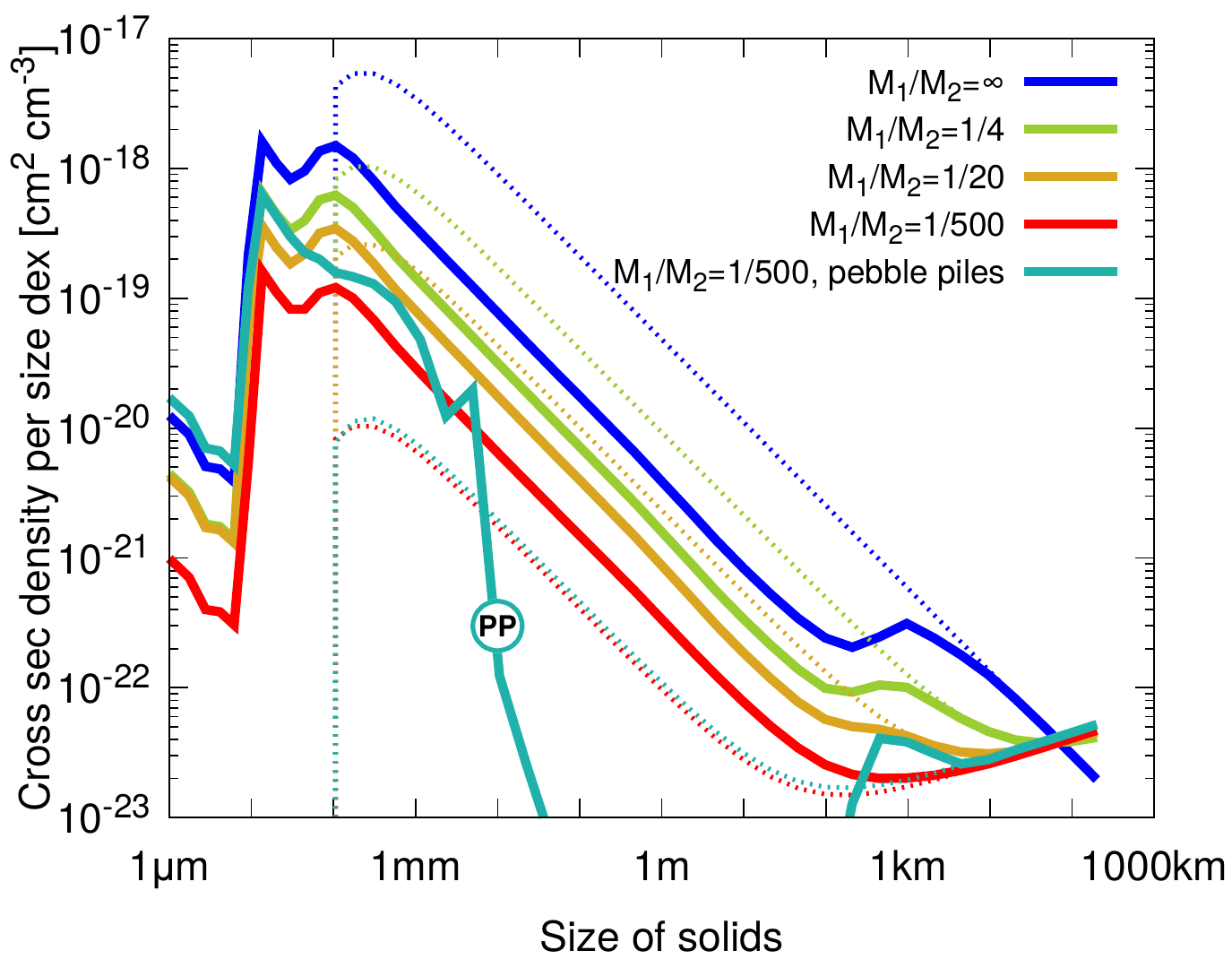}
\caption{Initial (thin dotted) and evolved (after $100\Myr$; thick solid)
size distributions of material in several fiducial debris discs,
produced by ACE simulations.
Different runs assumed the same total disc mass of $100 M_\oplus$, but different mass ratios 
between the two populations.
One of the curves (cyan, labelled ``PP'')
is the $M_1/M_2 = 1/500$ model assuming the alternative $Q_\text{D}^\ast$ model appropriate for pebble 
piles.
\label{fig:size_dist}
   }
\end{figure}

Figure~\ref{fig:size_dist} plots the size distribution of solids for some of the ACE runs.
Both the initial distribution and an evolved one 
after $100\Myr$ of collisional grinding
are depicted. The initial size distribution is a sum of the two distributions sketched in 
Fig.~\ref{fig:schematic}. Note that Fig.~\ref{fig:size_dist} plots the cross section surface density
per unit size decade,  whereas Fig.~\ref{fig:schematic} depicts the number of solids per unit decade in mass.
This explains a somewhat different appearance of the two figures. For instance, horizontal lines in 
Figure~\ref{fig:size_dist} (i.e., lines with a zero slope) would correspond to a slope $\alpha=5/3$
in Fig.~\ref{fig:schematic}. 

We start with discussing the initial distributions.
We see that low $M_1/M_2$ (or $s_\text{t}$) cases have an initial deficit of small grains 
(which are the ones that are seen by mid-IR surveys). Such discs need time to become bright.
This is  the time it takes to erode enough bigger objects in the initial size distribution 
in order to refill the initially missing dust population. Such a refill occurs because 
collisional cascade produces fragments following an $\alpha \approx 1.9$ size distribution that is steeper 
(and thus more small-dust ``friendly'') than the initial $\alpha=1.6$ one.

Further, Fig.~\ref{fig:size_dist} illustrates how evolved distributions of solids look like.
Common to all of the curves is a sharp cutoff at several micrometers, which is a classical 
radiation pressure blowout limit \citep{burns-et-al-1979}.
The grains on the left of it are nearly absent, as these are placed by radiation pressure
in hyperbolic orbits and leave the disc on dynamical timescales.
Above the blowout limit, the size distributions develop some ripples.
This ``waviness'' is also well known 
\citep{campo-et-al-1994b,krivov-et-al-2006,thebault-augereau-2007} and is a natural 
consequence of a sharp blowout cutoff and of a mass loss rate being
independent of size in logarithmic bins \citep{wyatt-et-al-2011}.
Going to larger sizes, a local minimum appears at 100s of meters for ``monolithic'' curves, which simply
mirrors the minimum of the critical fragmentation energy (see. Eq.~\ref{eq:QD}).

This ``mirroring'' of $Q_\text{D}^\ast$ by the size distribution is best illustrated with 
the alternative $Q_\text{D}^\ast$ model. Figure~\ref{fig:size_dist} shows how
a significant difference between the two models for the critical fragmentation energy (\ref{eq:QD}) and 
(\ref{eq:QD_pp}) causes a dramatic difference in the evolved size distribution.
This result is not surprising, as even the moderate changes in the critical fragmentation energy have been 
shown to alter the size distribution markedly \citep[e.g.][]{thebault-augereau-2007,gaspar-et-al-2012}. 
However, it is for the first time that such an extreme model for $Q_\text{D}^\ast$ is being studied.
The cyan curve, produced with the alternative 
$Q_\text{D}^\ast$ model, develops a pronounced minimum at sizes where $Q_\text{D}^\ast$
itself has a deep minimum.
The means that the disc contains a low number of
pebble piles that are small enough not to be kept together by gravity.

\begin{figure}
\centering
\includegraphics[width=0.49\textwidth, angle=0]{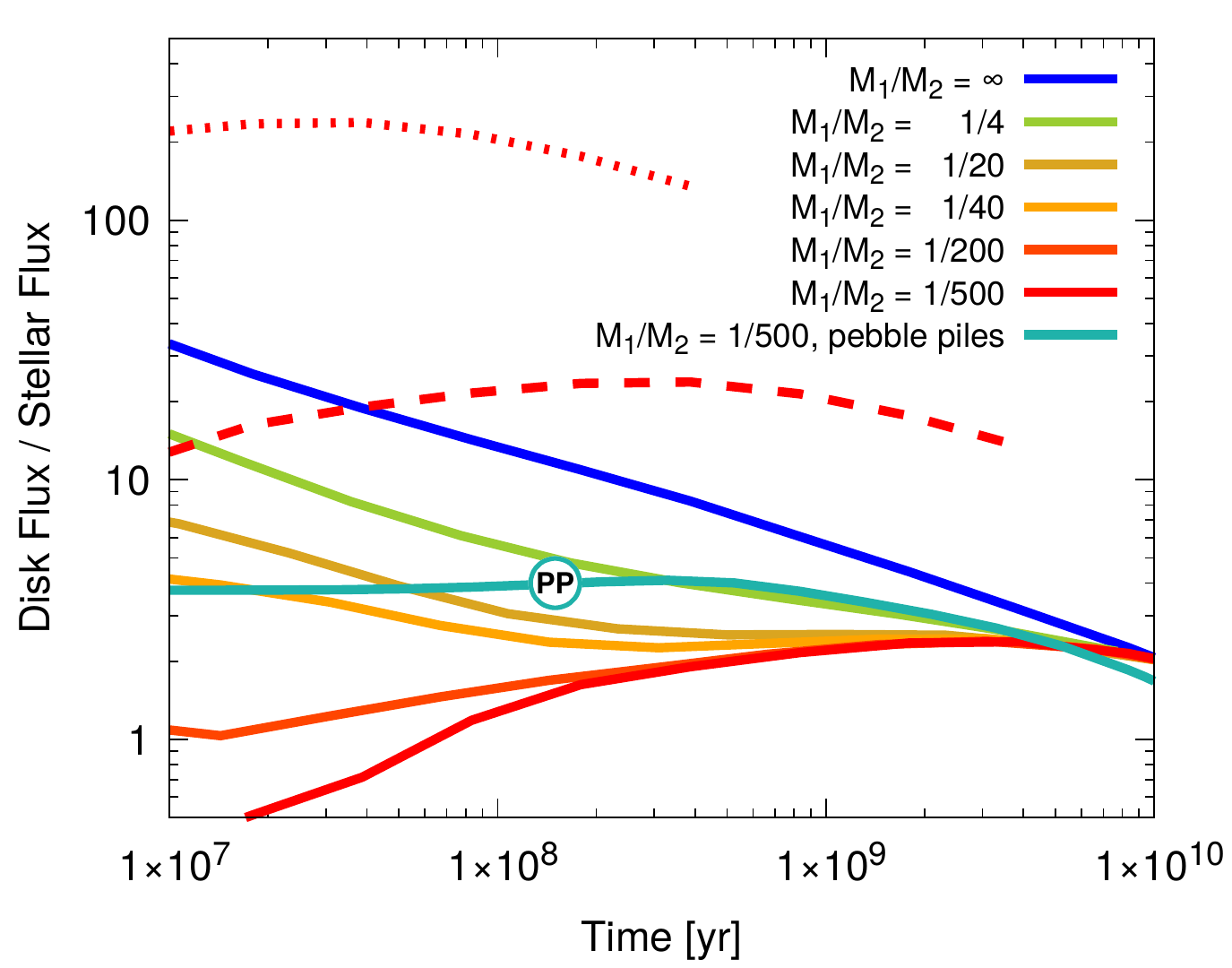}
\caption{Time evolution of a $70\mum$ flux from several fiducial debris discs, resulting from ACE 
simulations. 
The colour coding is the same as in Fig.~\ref{fig:size_dist}.
For the  $M_1/M_2 = 1/500$ model, the dashed and dotted lines show the flux 
evolution for discs with $1000 M_\oplus$
and $10,000 M_\oplus$ respectively. This illustrates the scaling described in the text, 
see Eq.~(\ref{eq:scaling}).
\label{fig:constant mass}
   }
\end{figure}

For each of the runs, we have also computed a time evolution of the dust emission flux at 
$70\mum$.
Figure~\ref{fig:constant mass} shows the results.
For the standard fragmentation energy model, the disc brightness decreases from higher to smaller
$M_1/M_2$ ratios. At the same time, the flux decay slows down. In fact, for a very pronounced population~2
the disc brightness first goes up until the decay sets in (see the $M_1/M_2 = 1/200$ and $M_1/M_2=1/500$
curves).
All this is because discs with a heavy population~2 contain nearly all their mass
in the large planetesimals, and there is initially not enough dust to emit.
That dust has first to be produced collisionally, which takes time.

This is different in the run that assumed an alternative $Q_\text{D}^\ast$ model.
In this case, planetesimals are fragile pebble piles that are collisionally destroyed very quickly.
This is because the pebble piles can be disrupted by much smaller projectiles than the monolithic 
planetesimals of the same size, and the collisional rates with smaller projectiles are much higher.
A debris disc consisting of such fragile planetesimals produces enough dust promptly, even if
population~1 is absent. This is demonstrated by a cyan curve in Figure~\ref{fig:constant mass}
that presents the $M_1/M_2=1/500$ model for the alternative $Q_\text{D}^\ast$ prescription.

While Fig.~\ref{fig:constant mass} presents the luminosity evolution of discs of the same total mass, 
it is also important to understand how a change in the total mass would alter the evolution.
To interpret the time evolution of discs with different initial masses, there is a useful 
analytic scaling \citep{loehne-et-al-2007,krivov-et-al-2008}. Consider a disc with initial 
mass $M_0$ at a distance $r$ from the star with age $t$.
Denote by $F(M_0,r,t)$ any quantity directly proportional
to the amount of disc material in any size regime, from dust grains to
planetesimals.
For instance, $F$ may stand for the total disc mass,
the mass of dust, its total cross section, or dust emission flux at any wavelength.
The scaling rule reads:
\be
       F(x M_0, r, t) = x F(M_0, r, x t) ,
\label{eq:scaling}
\ee
valid for any factor $x > 0$. Equation (\ref{eq:scaling}) holds for
every disc of particles, as long as these are produced, modified and lost in
binary collisions and not in any other physical processes.
Figure~\ref{fig:constant mass} illustrates, for one of the models, how the scaling works. 
When the initial mass of the disc increases, the flux evolution curve moves leftward and 
upward without changing its shape.

\section{Observational data}

To compare the collisional models with data, we chose to consider A-type stars.
There are more detections around A-type stars, providing better statistics.
Besides, the discs around A-type stars are known to exhibit a more rapid decay
of the dust luminosity towards older systems than discs of solar-type stars \citep{wyatt-2008}.
We selected three datasets:
\begin{itemize}
\item
a Spitzer/MIPS $70\mum$ sample of A-type stars from \citet{su-et-al-2006} that contains
69 excess stars with $S/N > 3$ (out of 137 in total) with ages between $5$ and $850\Myr$;
\item
a collection of Spitzer/MIPS $70\mum$ data
for different-type stars of \citet{chen-et-al-2014}.
From this, we extracted A-stars only (145 detections out of 149 observed A-type,
$3$ to $1360\Myr$-old stars);
\item
a Herschel/PACS $100\mum$ sample for A-type stars from \citet{thureau-et-al-2014}.
This includes 18 sources with a flux excess significance $>3$
among 83 stars with ages from $12$ to $800\Myr$.
\end{itemize}

\begin{figure*}
\centering
\includegraphics[width=0.82\textwidth, angle=0]{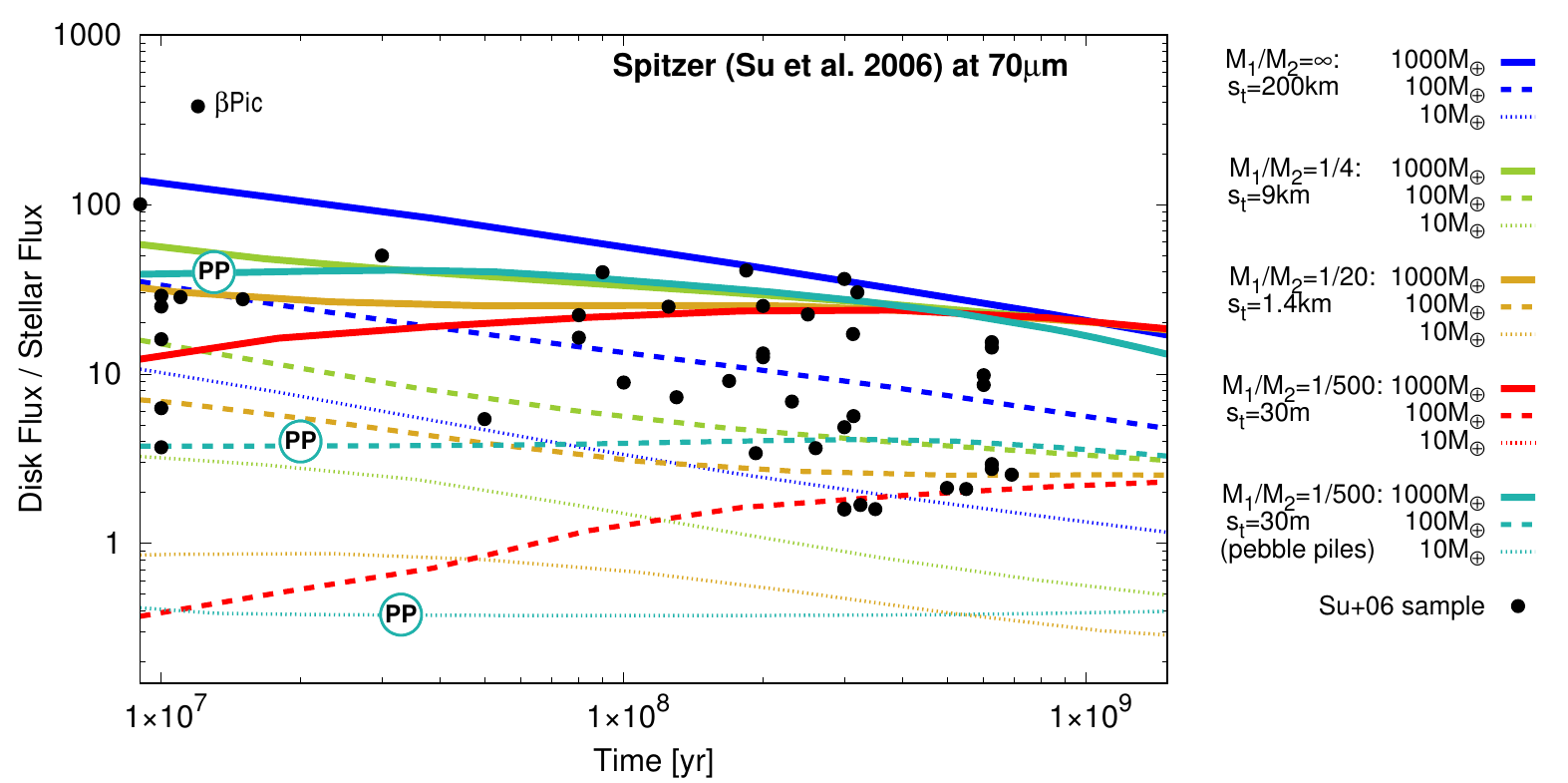}\\[-0mm]
\includegraphics[width=0.82\textwidth, angle=0]{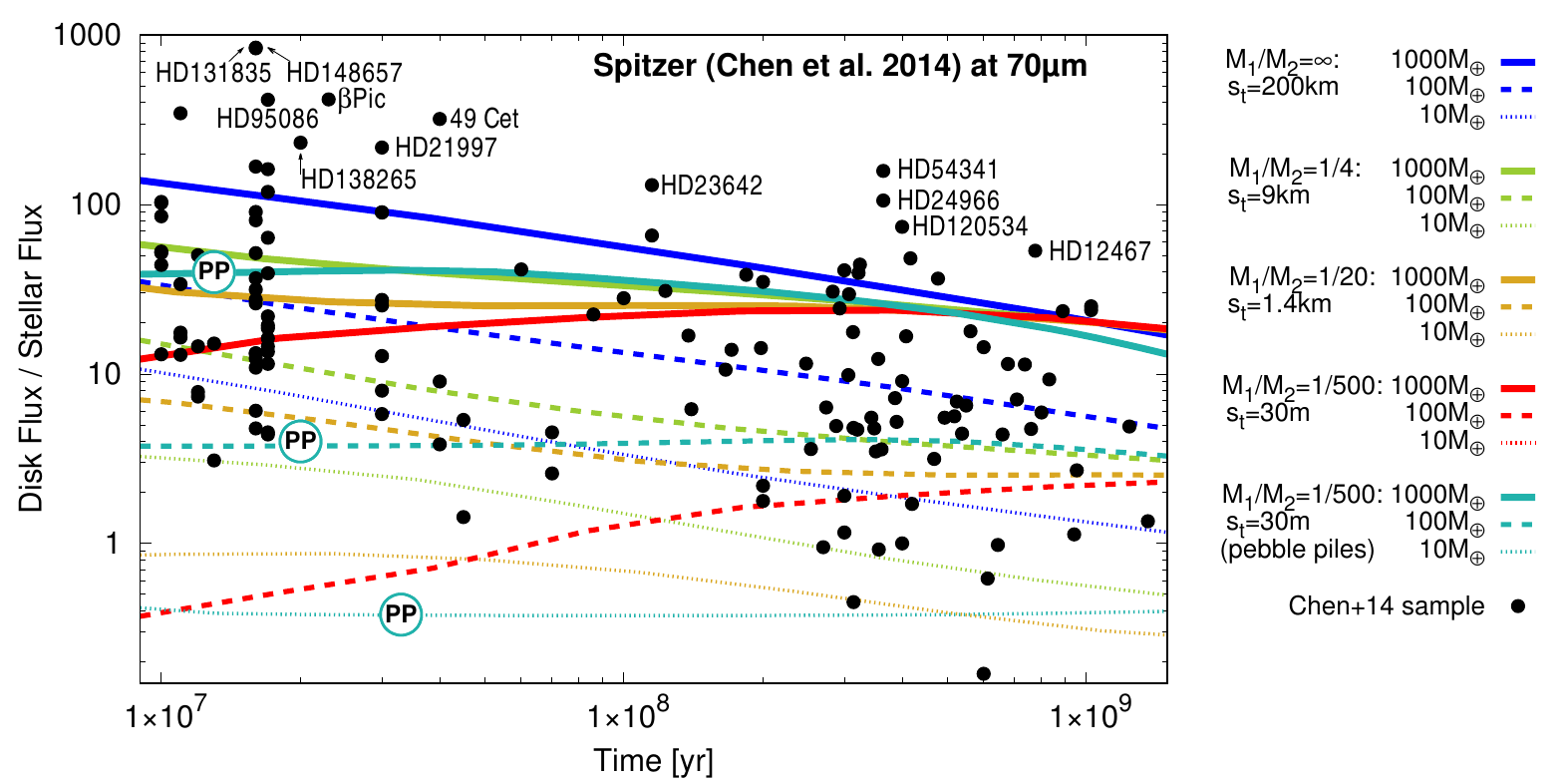}\\[-0mm]
\includegraphics[width=0.82\textwidth, angle=0]{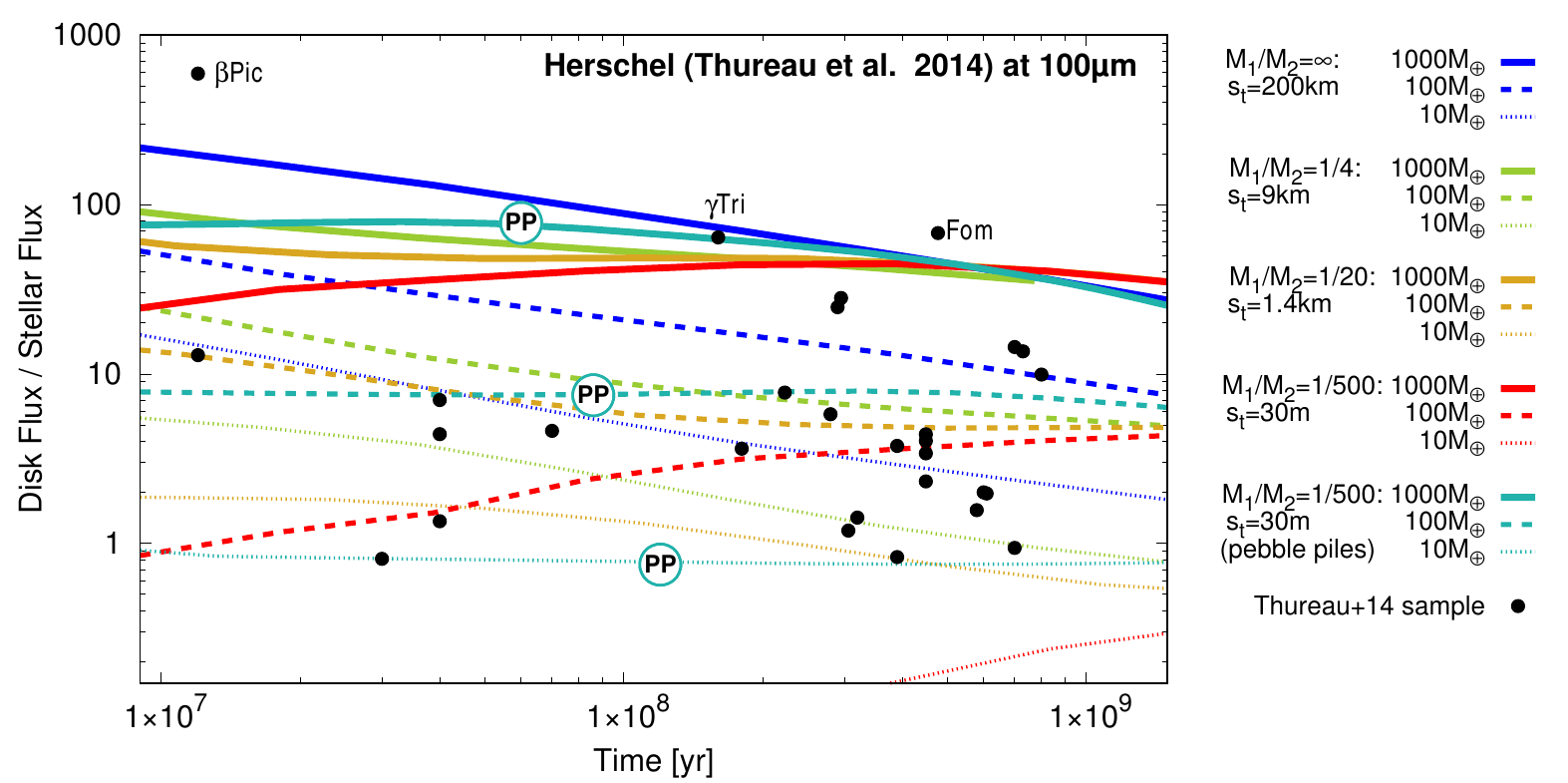}
\caption{
Samples from \citet{su-et-al-2006} (top),
\citet{chen-et-al-2014} (middle),
and \citet{thureau-et-al-2014} (bottom)
and the ACE models.
Symbols are flux ratios inferred from observations through the SED fitting, as published in the original papers.
A few prominent discs are labelled.
Lines are ACE models (different colours and linestyles for different $M_1/M_2$,
and different line thicknesses for different $M_1+M_2$).  For each $M_1/M_2$, three curves for $M_1+M_2$ of 10, 100, and 1000
Earth masses are shown.
As in previous figures, the cyan curves labelled ``PP'' correspond to 
$M_1/M_2 = 1/500$ and the alternative  $Q_\text{D}^\ast$.
\label{fig:Comparison with samples}
}
\end{figure*}

These three samples are visualized in Fig. \ref{fig:Comparison with samples}. It plots
the excess ratios (i.e., dust fluxes divided by the stellar photospheric flux) versus stellar age.
Before we proceed with a comparison between these data and collisional models, we make several remarks about the
data themselves.

First, \citet{chen-et-al-2014} and \citet{thureau-et-al-2014} also performed the SED fitting of 
the excess stars. They tried both single- and two-temperature fits and concluded that many of the
discs are 
likely to have a two-component structure, including an additional warm component closer to the star.
Since we discuss the cold Kuiper-belt analogs in this paper, it would be reasonable to subtract the 
contribution of the warm component from the total disc flux.
We now estimate this contribution. Denote the temperatures of the warm and cold components
by $T_w$ and $T_c$, and their fractional luminosities by 
$f_w$ and $f_c$, respectively. Next, let $R_w(\lambda)$ and $R_c(\lambda)$ be the excess ratios
for the two components.
Assuming both components to emit as black bodies, it is easy to derive
\be
{R_w (\lambda) \over R_c (\lambda)}
=
\left(f_w \over f_c \right)
\left(T_w \over T_c \right)^{-4}
\left[B_\nu (T_w, \lambda) \over B_\nu (T_c, \lambda) \right] ,
\label{eq:flux ratio}
\ee
where
$$
B_\nu (T, \lambda)
=
{2 h c\over \lambda^3}
\left[ \exp\left( h c \over \lambda k T \right) -1 \right]^{-1} 
$$
is the Planck intensity in the frequency scale as a function of temperature and wavelength,
with $h$ being Planck constant, $c$ the speed of light, and $k$ the Boltzmann constant.
For the  \citet{chen-et-al-2014} sample, the average values are
$T_w=300\K$, $T_c=100\K$,
$f_w=6\times 10^{-5}$, and $f_w=1\times 10^{-4}$.
Equation~(\ref{eq:flux ratio}) gives $R_w/R_c$ as small 2.9\% at $70\mum$ and 2.2\% at $100\mum$,
showing that the contribution of the inner component to the observed flux
at $70\mum$ and $100\mum$ is almost negligible.
We decided against making this correction for the sake of simplicity.

Second, there are some caveats about the datasets that are related to the brightest 
discs with excess ratios of $\ga 100$, which are the most numerous in the
\citet{chen-et-al-2014} sample.
These discs can be loosely classified into two groups, the old
($> 100\Myr$) and the young ones ($< 100\Myr$), and we now discuss these two groups in turn.
The brightest discs around stars listed as having ages $100\Myr$ and more in
their sample, most of which do not belong to clusters, may in fact be considerably younger 
according to independent age estimates reported in the literature. For example, HD~54341 has 
a reported age of $364\Myr$ in \citet{chen-et-al-2014},
while an age estimation based on isochrones suggest a much younger age of $10\Myr$ 
\citep{rhee-et-al-2007}. 
The same applies to several other bright discs reported by \citet{chen-et-al-2014} to be old
~--- they all may be much younger. 
These include
HD~12467 \citep[$200\Myr$ instead of $776\Myr$, see][]{rhee-et-al-2007},
HD~24966 \citep[$100\Myr$ instead of $364\Myr$, see][]{moor-et-al-2006},
and HD~120534 \citep[$<320\Myr$ instead of $400\Myr$, see][]{moor-et-al-2006}.
This would push the bright but old discs to much younger ages, where high dust 
luminosities are readily expected. 
For bright discs around stars younger than $100\Myr$,
it is likely  that the ``collisional'' ages of our numerical model do not necessarily correspond to
the system ages. Some discs could have started the collisional cascade later in their lifetimes due to 
delayed stirring; we will discuss this in Section~5.3.
Finally, in many of the youngest, bright discs that fill the left upper corner
in the middle panel of Fig. \ref{fig:Comparison with samples},
gas has been detected \citep[e.g.,][]{moor-et-al-2011,kral-et-al-2017c}.
There is an uncertainty of whether those are true 
debris discs or are still in the protoplanetary or transitional phase.
Our model would not be applicable to such systems.

Another star with a bright disc in the Pleiades (age of $\approx 100\Myr$) in the 
\citet{chen-et-al-2014}  sample, HD~23642, is an eclipsing binary with a period of about 
2.5 days \citep[see][where also the stellar parameters are given]{southworth-et-al-2005}.
Fitting a model photosphere of a single star to the binary star
will underestimate the IR luminosity of the binary, if the smaller companion is colder 
than the primary one.
This happens because the bright and hot star dominates the flux at short wavelengths 
where the fitting takes place, but the influence
of the companion grows at the longer wavelengths, meaning the flux ratio of secondary to 
primary star, $F_\text{B}(\lambda)/F_\text{A}(\lambda)$, is larger in far-IR.
For HD 23642, this flux ratio is about $30$\% higher at $70\mum$ than at 
$1\mum$.
Taking this effect into account reduces the flux ratio for the disc of HD 23642
from 130 as reported by \citet{chen-et-al-2014} to
$130 \times (1+F_\text{B}(1\mum)/F_\text{A}(1\mum))/(1+F_\text{B}(70\mum)/F_\text{A}(70\mum))\approx 125$.
Although the correction is moderate, it may affect a number of 
discs,
since some of their host stars may have as yet unknown colder companions.
In addition, close binarity may directly influence the discs
by perturbing and stirring them.

 
\section{Model vs data}

Figure \ref{fig:Comparison with samples} compares the \citet{su-et-al-2006},
\citet{chen-et-al-2014}, and \citet{thureau-et-al-2014} data with our collisional model
for different choices of $M_1+M_2$ and $M_1/M_2$. 
All of the simulations but one assumed the standard 
$Q_\text{D}^\ast$ model
(Eq.~\ref{eq:QD}). One simulation was run with the alternative $Q_\text{D}^\ast$
assuming pebble piles (Eq.~\ref{eq:QD_pp}).
In each case, three curves are presented, corresponding
to the $M_1+M_2=10$, $100$, and $1000 M_\oplus$.

\subsection{Discs of ``monolithic'' planetesimals}

We start with a discussion of the runs done with a standard $Q_\text{D}^\ast$
(i.e., ``monolithic'' planetesimals).
We compare the simulations with the data in the order of decreasing $M_1/M_2$
(or increasing $M_2$, or decreasing $s_\text{t}$). Purely ``traditional'' models, i.e.,
those without population~2, reproduce both the mean emission level and its long-term 
decay with the stellar age at disc masses of $10$--$100 M_\oplus$.
This is consistent with \citet{wyatt-et-al-2007b} who derive
a mean disc mass of $10 M_\oplus$ (assuming $s_\text{max}$ of $30\km$) in their population model
designed to reproduce the sample of discs around 
A-type stars from \citet{rieke-et-al-2005}.
Adding the ``non-traditional'' population~2 with $s_\text{t}$ at several kilometers,
we are still able to reproduce the data, as expected
(see Fig.~\ref{fig:constant mass} and a discussion of it). However,
the disc masses $M_1+M_2$ needed to reach the observed emission level increase.
Between almost a hundred and a few hundred of Earth masses are now needed to match the
discs of average brightness.

Decreasing $s_\text{t}$ to about one kilometer leads to growing discrepancies between the data and 
models.
The decay slows down.
For $s_\text{t} \sim 1\km$ (see the $M_1/M_2=1/20$, or $s_\text{t}=1.4\km$ curves), the flux evolution
gets nearly flat.
For yet smaller $s_\text{t}$, the disc brightness evolution is no longer monotonic.
Instead, it first increases for at least $100\Myr$ and only 
then starts to decrease
(see $M_1/M_2 = 1/500$ or $s_\text{t}= 30\m$ curves), which is inconsistent with the data.
Even though the scatter of points at any given age is large, it is obvious
that the models that only include population~2 do not provide any 
acceptable fits to the observed trend.
This tells us that,
for the initial size distribution of planetesimals to be consistent with the data,
their size distribution should steepen at sizes below about one kilometer.
In other words,
a protoplanetary disc that has successfully built planetesimals with 
a mass distribution slope of $\alpha = 1.6$ must also contain a substantial amount of smaller,
sub-kilometer-sized solids
from the very beginning. 

\begin{figure}
\centering
\includegraphics[width=0.49\textwidth, angle=0]{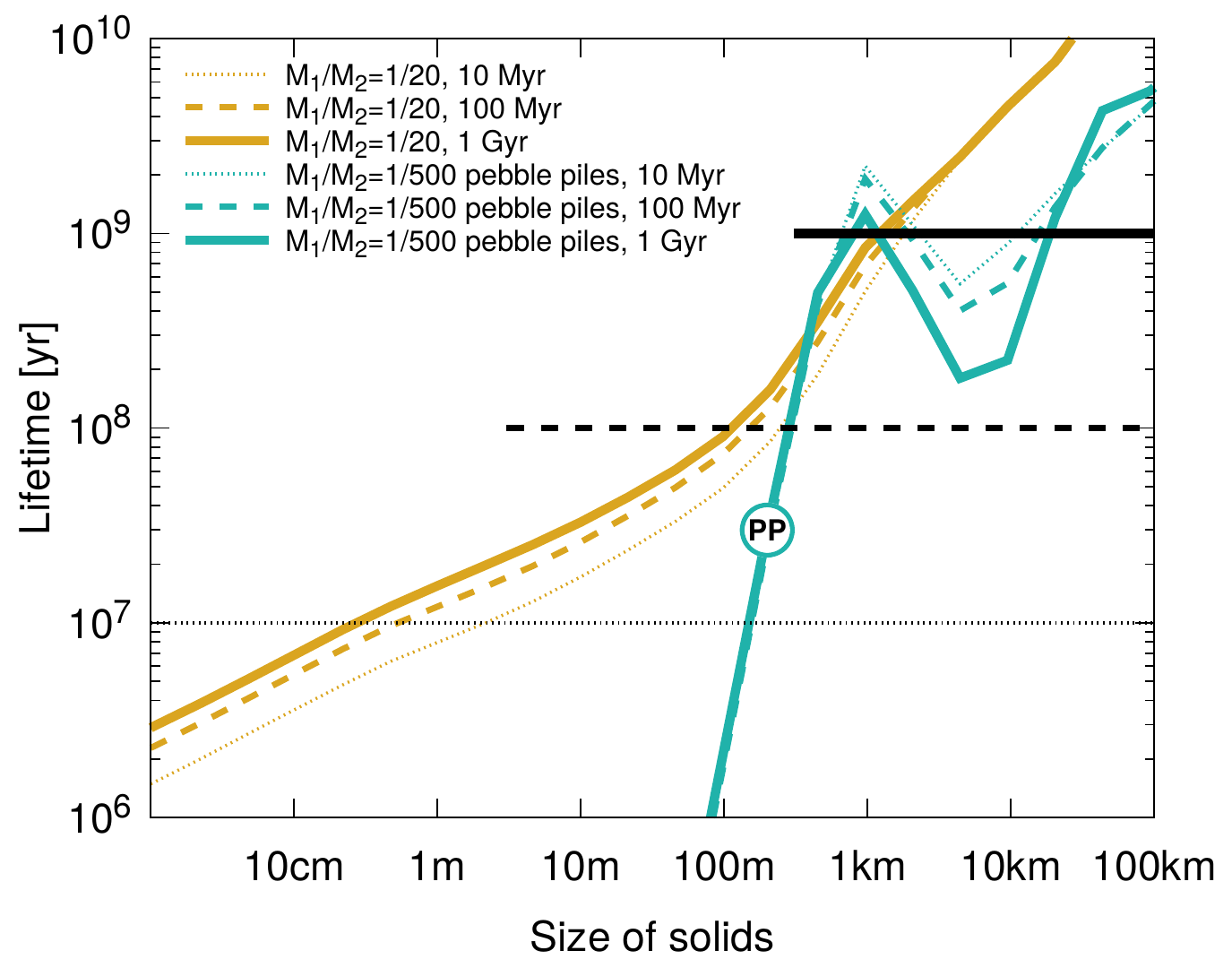}\\
\caption{
Collisional lifetimes of different-sized solids in two ACE runs (yellow and cyan, same colours
as in previous figures) after $10\Myr$ (thin dotted), $100\Myr$ (medium dashed) and $1\Gyr$ (think solid)
of collisional evolution.
These three characteristic ages themselves are shown with horizontal straight lines
of the same linestyles.
\label{fig:Tcoll}
   }
\end{figure}

What is the minimum size of those ``additional'' objects?
At any time of the collisional evolution, they should be large enough to not get
collisionally depleted by that time.
Figure~\ref{fig:Tcoll} plots collisional lifetimes of different-sized objects in one of the ACE runs
discussed above. Yellow curves present the
$M_1/M_2=1/20$ (or $s_\text{t}=1.4\km$) model for monolithic planetesimals which~--- as shown above~---
reproduces the observed emission
at all ages from $\sim 10\Myr$ to $\sim 1\Gyr$ reasonably well. It is seen that
the objects should be larger than at least a few meters in size to survive the first $10\Myr$.
At $100\Myr$,
the minimum radius should be about $100\m$. At $1\Gyr$, it grows to at least one kilometer.
We can conclude that, to sustain the observed amount of emission, the disc must contain
solids in the size range from $\sim 1\m$ to $\sim 1\km$, with a size distribution close to
Dohnanyi's ($\alpha \approx 1.8...1.9$).

In a standard, ``slow-growth'' scenario, such objects indeed appear as a natural aftermath of formation
of large bodies
\citep[e.g.,][]{kenyon-bromley-2008,kobayashi-et-al-2010b,kobayashi-loehne-2014,%
kobayashi-et-al-2016}.
As explained in Section 2.4, growth of larger bodies in this scenario keeps pace
with fragmentation at smaller sizes. 
At early stages, when a big amount of gas is still present in a protoplanetary disc, the 
collisional cascade is stalled at around 1--10 meter sizes, because the relative velocities of 
smaller objects are damped by the ambient gas.
However, once gas depletion becomes significant, collisional cascade promptly extends to dust sizes.
As a result, by the time of the gas dispersal one expects a Dohnanyi-like slope
at sizes below about a kilometer and a shallower slope for larger bodies that have grown in the disc
(H. Kobayashi, pers. comm.).

\subsection{Discs of ``pebble-pile'' planetesimals}

We now turn to a discussion of the case where planetesimals are assumed to be pebble piles and are described
by the alternative $Q_\text{D}^\ast$. The cyan curves in Fig. \ref{fig:Comparison with samples} demonstrate
that the simulated brightness evolution is not inconsistent with the sample, even if population~1 is nearly absent
(we took $M_1/M_2=1/500$).
In contrast to monolithic objects, pebble piles
generate a considerable amount of dust early on.
As discussed above, this is because even collisions with much smaller projeciles, which are quite
frequent, are sufficient to disrupt pebble piles. 
As a result, their collisional lifetimes are very short,
dropping to below
$1~\Myr$ for planetesimals smaller than 100 meters in radius
(see cyan curve in Fig.~\ref{fig:Tcoll}).

Still, the dust brightness curves in Fig. \ref{fig:Comparison with samples}
might be too flat, not matching closely a gentle long-term decay that is best 
visible in the \citet{chen-et-al-2014} sample.
However, the underlying prescription~(\ref{eq:QD_pp}) of $Q_\text{D}^\ast$ 
used in the simulation should only be considered as a rough proxy of how the strength
of ``pebble piles'' may look like in reality.
If the tensile strength of dust is higher than in our model, of if the 
minimum of $Q_D^\star$ at meter sizes is deeper than
in Eq.~(\ref{eq:A_pp}), then the discs will be brighter at young ages (see a discussion in 
Section~5.4).
Also, assuming smaller pebbles~--- e.g., $1\mm$ instead of $1\cm$~--- would help, since shorter collisional 
lifetimes of tinier pebbles would enhance dust production early on. In fact, pebbles of a millimeter or 
even  smaller sizes can be expected in outer zones of the discs
\citep[see, e.g., Fig. 3a in][]{johansen-et-al-2014}.
This is because the particle sizes that correspond to the ``right'' Stokes numbers St $\sim 0.01$--$0.1$ (at 
which concentration by streaming instability is efficient) decrease at larger distances from the star
\citep[e.g.][]{carrera-et-al-2017}.
This or that way, it is likely that more realistic models would provide a better match to the observed 
evolution.

The low fragmentation energy of pebble piles also implies
that the disc is devoid of small, sub-km-sized
planetesimals (see cyan curve in Fig.~\ref{fig:size_dist}).
This prediction is potentially testable with the Solar System objects. We will
discuss this in more detail in Section~5.1.

\subsection{Total masses of bright debris discs}

An inspection of Fig.~\ref{fig:Comparison with samples} uncovers a problem that may have been overlooked previously.
It concerns the brighter discs, especially in the \citet{chen-et-al-2014} sample. Even if we discard 
the brightest outliers, as discussed in Section~3, the total disc mass required to reproduce bright discs
with the models is as large as $\sim 1000 M_\oplus$.
This is true for both monolithic and pebble-pile $Q_\text{D}^\ast$ prescriptions. 

The required total mass of the disc increases with decreasing $M_1/M_2$ ratio,
i.e., if a heavier population~2 is added.
However, large total disc masses have already appeared in previous works, based on
``traditional'' models without population~2.
For example,
\citet{wyatt-dent-2002} require at $20$--$30 M_\oplus$ in bodies with radius of less than $2\km$,
which would translate to $\sim 200$--$300 M_\oplus$  in $200\km$-sized objects, to explain the disc of Fomalhaut.
The ``high-mass discs'' with a dust mass of 0.1 $M_\oplus$
investigated in \citet{thebault-augereau-2007} imply $190$--$850 M_\oplus$ in planetesimals of up to $50\km$ in radius
(see their Table~2).
\citet{mueller-et-al-2009} derive $50$--$60 M_\oplus$ in
bodies smaller than $74\km$ for the Vega disc. 
\citet{kobayashi-loehne-2014} reproduce the evolution of discs around solar-type stars
with $2\AU$-wide rings with a radius of $30\AU$, containing $45 M_\oplus$ in $100\km$-sized 
planetesimals. \citet{schueppler-et-al-2016}
obtain $165 M_\oplus$ for bodies up to $50\km$ in size to explain the disc
around $q^1$~Eridani.

How massive is an outer planetesimal belt actually allowed to be?
Let us consider an initial protoplanetary disc (PPD), which is a progenitor to the debris disc in question.
Obviously, it should be lighter than the central star, so let us assume the maximum initial PPD mass
is $0.1 M_\star$. For simplicity, let the central star
be sun-like. Assuming a standard gas-to-dust ratio of 100:1, we
conclude that the maximum total mass in solids in the entire disc is
$0.001 M_\odot$, or $300 M_\oplus$. By going to more massive stars (e.g.
A-stars with 2 or 3 solar masses) or by assuming somewhat lower gas-to-dust
ratios \citep[such as the Solar System's abundances giving $60:1$, see][]{lodders-2003},
we can obviously increase that total mass in solids by a factor of two or so,
but not much more. This initial constraint is robust. There are simply not
more solids available in the PPD than that.
This is consistent with the masses of the observed protoplanetary discs.
For instance, Fig. 5 in \citet{williams-cieza-2011} suggests the total
masses of PPDs around A-stars to be between 2 and 200
Jupiter masses or $600$--$60 000M_\oplus$, implying $6$--$600 M_\oplus$ in solids.

\begin{figure}
\centering
\includegraphics[width=0.49\textwidth, angle=0]{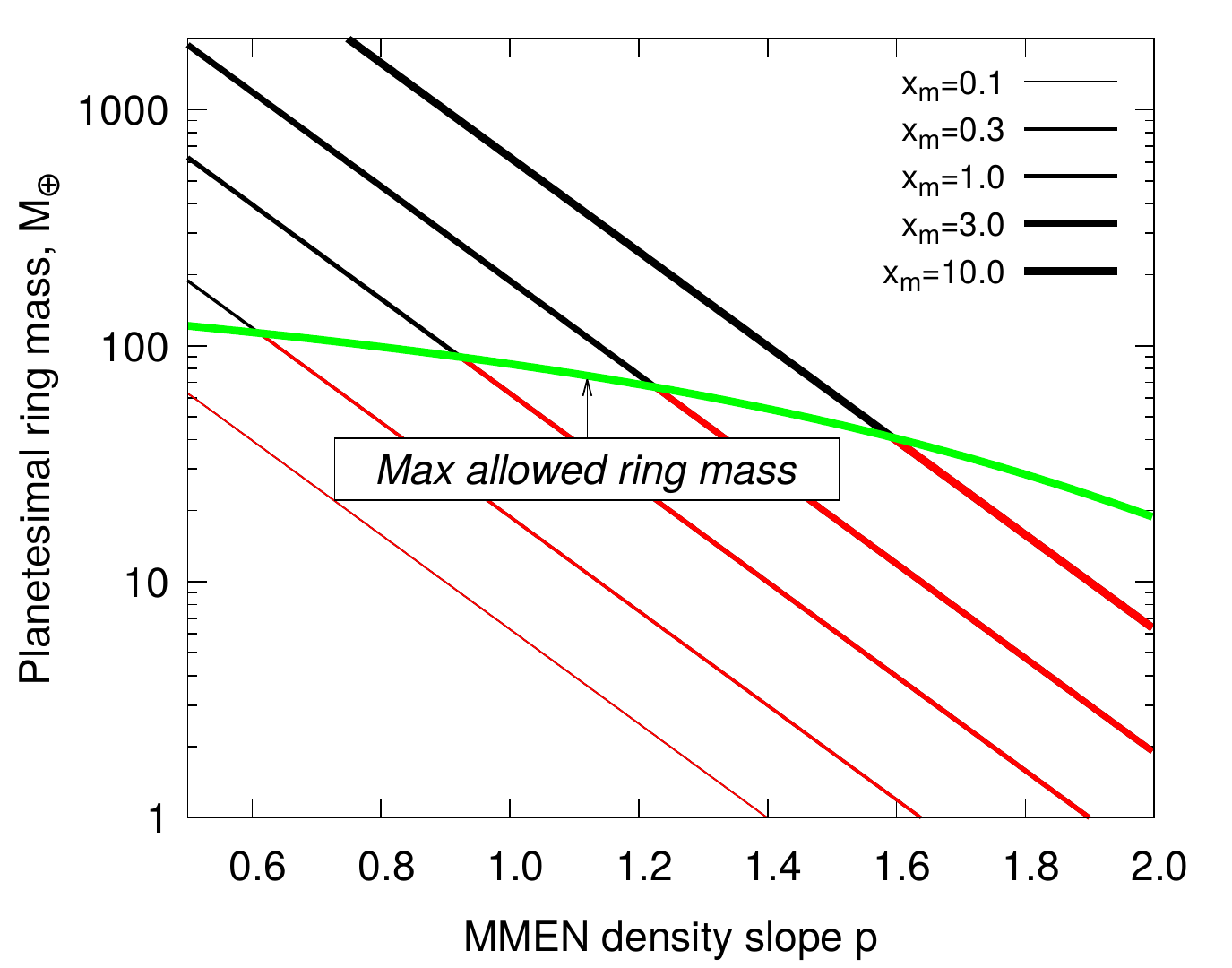}\\
\caption{
The masses of $10\AU$-wide planetesimal rings at a distance of $100\AU$ from the
A0-star of $2.9M_\odot$ in the MMEN model.
Solid lines are ring masses as functions of $p$ (from thin to thick:
$x_m = 0.1$, $0.3$, $1$, $3$, and $10$).
Red portions of these lines correspond to rings such that the total mass of the
disc of solids (if the ring cavity were absent) does not exceed 0.001 of the stellar mass.
Black portions mean it does; such rings are not allowed.
The green line separating red and black represents the maximum allowed
ring masses.
\label{fig:MMEN}
}
\end{figure}

\citet{kuchner-2004} introduced the term ``Minimum-Mass Extrasolar Nebula'' (MMEN)
for extrasolar planetary systems by analogy to the
Minimum-Mass Solar Nebula \citep{weidenschilling-1977b,hayashi-1981} for our Solar System.
The MMEN concept parameterizes the surface density distribution in a PPD by two parameters,
the density scaling $x_m$ and the radial slope $p$
(such that MMSN has $x_m=1$ and $p=1.5$).
Using MMEN, we can easily estimate the solid mass sitting in the
outer zone of a certain radius (e.g., $100\AU$) and a certain width (e.g., $10\AU$).
The result (green line in Fig.~\ref{fig:MMEN}) is
$20$--$130 M_\oplus$ for discs of A0-stars.

This estimate tacitly assumed that there was no substantial
radial transport of solids at earlier phases. However,
radial drift in combination with other processes acting  
at the planetesimal formation phase
displaces the solids  radially. As a result, the density profile of planetesimals
differs from the initial density profile of solids \citep{carrera-et-al-2017}.
Another possibility is scattering the
planetesimals outward by giant planets later on. This is similar to the
``implanted Kuiper belt'' scenario of \citet{levison-et-al-2008}.
The mass in giant planets needed for such a displacement is moderate.
For instance, to displace $100 M_\oplus$
outward one would need about the same mass in giants, so one Saturn or six Neptunes would do.
A caveat is that most of the planetesimals will be ejected to
interstellar space, and some will fall down to the star.
Only a moderate fraction
will be implanted to the outer zone we are talking about.

The estimates above are also consistent with predictions of planetesimal formation 
simulations. Assuming a PPD of $0.1M_\star$ with $p=1$, \citet{carrera-et-al-2017} robustly produce
a total of $80$--$150 M_\oplus$ in planetesimals, of which
a $60$--$130 M_\oplus$ are located outside $100\AU$.

In summary, several $100 M_\oplus$
is the absolute upper limit on the planetesimal disc mass, whereas collisional models~-- both 
previous ones and those considered here~--- often require much larger masses.
We will refer to this as a ``disc mass problem'' and discuss possible solutions to it in Section~5.

\section{Discussion}

A comparison of models to the debris disc samples has led us to two important findings.
One is that debris disc evolution can be successfully reproduced both with the slow-growth and particle 
concentration scenarios of planetesimal formation. Can we distinguish between the two?
Another one
is what we call ``the disc mass problem.''
We have found a discrepancy between the planetesimal disc masses 
expected from their formation models (likely $\sim 100 M_\oplus$, but not more than a few 100s $M_\oplus$)
and masses needed to reproduce the observed emission of the brightest debris discs
(up to $1000 M_\oplus$ in rings of $10\AU$ in width).
Is it possible to make the total mass of a
planetesimal belt smaller, while keeping the same amount of visible dust?

We now try to answer these questions, the first one in Sect.~5.1 and the second one in Sect.~5.2--5.6.
As we will see, especially the disc mass problem may require 
addressing more fundamental aspects of debris discs,
whose importance goes beyond that particular problem.

\subsection{Slow growth or pebble concentration?}

Although we cannot discriminate between the two possible routes of planetesimal formation
from the analysis of dust fluxes, one possible way to distinguish between them
is through the drastically different predictions
they make for the amount of sub-km-sized bodies.
This is of little help for debris discs around other stars where such objects are not accessible,
but at least we may look at our Solar System where sub-km-sized bodies are directly observable.
In the present-day Kuiper belt, the spatial density of objects is so low that only the objects 
smaller than several tens of meters have collisional lifetimes shorter than the age of the Solar System
\citep{vitense-et-al-2012}.
In contrast to extrasolar debris 
discs, it is these sub-km-sized bodes in the Kuiper belt that are located at the top of the collisional 
cascade and serve as the actual source of dust. To be consistent with spacecraft 
dust detections in the outer Solar System, these bodies should be present in amounts much lower than what 
would be expected by Dohnanyi-like extrapolation from the observed, 10--$100\km$-sized 
transneptunian objects to sub-kilometer sizes.
Consistently, \citet{fulle-blum-2017} point out that the crater density observed on the bodies in the outer 
Solar System is low, suggesting that the number of comets smaller than a few kilometers is lower than 
assumed so far.
Closer to the Sun, the size distribution slope of asteroids in comet-like orbits flattens below about 
one kilometer in size \citep{kim-et-al-2014}.
Finally, \citet{trilling-et-al-2017} analyzed the size distribution of near-Earth asteroids from 1 km down to 
10 meters and found that there are
a factor of ten fewer small objects than assumed previously.
All this seems to speak for the suggested 
reduced amount of such bodies in debris discs and thus, to favour the pebble concentration scenario.
There is also more direct empirical evidence that comets in the Solar System
likely formed through the gentle gravitational collapse of a bound clump 
of mm-sized pebbles, intermixed with microscopic ice particles.
Using the results obtained with a suite of instruments on-board Rosetta mission to comet
67P/Churyumov-Gerasimenko, \citet{blum-et-al-2017} found this
formation scenario to be compatible with the measured global porosity, homogeneity, tensile strength, thermal 
inertia, vertical temperature profiles, sizes and porosities of emitted dust, and the water-vapour 
production rate of the comet.
The caveat is that we do not know how represenative 67P is for the entire cometary population,
and how respesentative the Solar System is for other planetary systems.

\subsection{No big planetesimals?}
At a first glance, a simple remedy to the disc mass problem would be to assume that the 
planetesimals larger than a kilometer in size are completely absent.
Indeed, it is known that to sustain the observed level of brightness for typical debris discs over Gyrs,
kilometer-sized planetesimals would suffice \citep[e.g.,][]{loehne-et-al-2011}.
This can also be seen  in our simulations.
Figure~\ref{fig:Tcoll} suggests that
lifetimes of all objects smaller than $1$--$10\km$ are typically shorter than systems' 
ages, so that their size distribution is set by the collisional cascade.
In contrast, the planetesimals larger than
$10\km$ in size essentially retain their primordial distribution.
They make a minor contribution to the
cascade, only producing a moderate amount of dust through cratering collisions.

Since the disc mass is proportional to $s_\text{max}^{6-3\alpha}$, replacing
$s_\text{max}=100\km$ with $s_\text{max}=1\km$ would reduce the belt mass by a factor of 5
for $\alpha = 1.88$
or even by a factor of 250 for $\alpha = 1.6$.
However, that would challenge the planetesimal formation models that robustly 
predict largest planetesimals to be 100s of km in size \citep{schaefer-et-al-2017}.
Furthermore, without large planetesimals it would be difficult to explain why the discs produce dust at all.
Dust is only produced if the planetesimals have sufficiently high relative velocities, i.e., non-zero 
eccentricities and/or inclinations.
Stirring may come from embedded big planetesimals or from planets in the inner cavities of the discs.
In the former case,
this directly requires large planetesimals to be present. In the latter case,
it would be difficult to understand why planetary-sized bodies have formed successfully closer in,
whereas 100km-sized planetesimals slightly farther out have not.

Even though the predicted slope of the size distribution of planetesimals probably extends to 
large sizes, it remains possible that the actual $s_\text{max}$ is by a factor of several smaller 
than assumed here. For instance, recent simulations by \citet{klahr-schreiber-2016} produce
planetesimals with sizes peaking at $\approx 50\km$. Such sizes would imply the required total 
masses of debris discs by a factor of 2--4 smaller, reducing the tension between the debris disc 
models and the data.

\subsection{More recent origin of debris?}
All the results obtained above rely on two basic assumptions.
One is an assumption that the observed debris dust is produced in a collisional 
cascade, which is treated statistically. This means that even the largest colliding bodies are 
considered to be numerous enough to be part of a continuous size distribution \citep{tanaka-nakazawa-1994}.
Another assumption is that the ``collisional age'' of the systems, i.e., the time elapsed since
the ignition of the cascade, is equal to the system's age. In principle, we can question both.
Especially for the brightest discs at older ages, it is possible that a recent giant impact
caused a temporary brightness increase 
\citep[e.g.][]{kenyon-bromley-2005,jackson-wyatt-2014,kral-et-al-2014,genda-et-al-2015}.
Also, ``delayed stirring''
\citep{dominik-decin-2003,wyatt-2008} is 
possible, implying that the cascade was triggered later in the system's 
history~--- for instance following the formation of Pluto-sized planetesimals in 
the disc \citep{kenyon-bromley-2008},
when the stirring front from planets in the inner cavities has reached the
disc \citep{mustill-wyatt-2009}, or by late dynamical instabilities caused by planetary migration or 
scattering \citep{booth-et-al-2009}.
In all these cases, the total mass of a debris disc of a given brightness would be lower than
derived in this work.

\subsection{Overall steeper size distribution?}
Another way to get more dust from a given total mass would be to assume
that the overall slope of the size distribution of solids
from dust to km-sized planetesimals is steeper than collisional simulations suggest.
However, that slope cannot simply be varied, as it is set by the cascade.
Specifically, it depends on the slope of $Q_D^\star (s)$  \citep{o'brien-greenberg-2003}.
This can easily be explained with simple qualitative arguments.
The amount of dust is determined by the product of the dust production rate and lifetime.
The production rate equals mass loss rate of planetesimals
(since the cascade relays the mass from planetesimals to dust),
while the dust lifetime is set by $Q_D^\star$
\citep[being proportional to $(Q_D^\star)^{5/6}$, see][]{wyatt-et-al-2007}.
In turn, the mass loss rate of planetesimals is set by their $Q_D^\star$.
So, effectively, the mass ratio between dust and planetesimals depends on the respective $Q_D^\star$ ratio.
Making the planetesimals at the top end of the cascade weaker
(which corresponds to steepening the $Q_D^\star$ slope)
should have a similar effect as directly strengthening the dust.
However, the $Q_D^\star$ of the biggest planetesimals is largely determined by gravity and thus is quite certain.
Thus the only way to increase the amount of visible dust is to assume that the dust is ``harder'', i.e.,
has a larger $Q_D^\star$.

The above reasoning also means that the 
$Q_D^\star$ of intermediate, centimeter- to kilometer-sized bodies should be unimportant
for the long-term evolution of discs.
This is readily confirmed by our simulations.
To see this, it is sufficient to compare the cyan curve in Fig.~\ref{fig:size_dist} to the other curves.
The cyan curve employed an alternative critical fragmentation energy model 
to describe the destruction of pebble-pile, ``macrogranular'' planetesimals,
while the others used a standard prescription.
Remember that the alternative model predicts much weaker planetesimals with radii between the pebble size and the sizes
where the strength starts to be dominated by gravity, but does not change much the strength of both the {\it dust}
and {\em large planetesimals}.
A comparison shows that all these simulations predict a comparable amount of dust after $100\Myr$ of evolution.
Note that this is only true after the disc has evolved long enough for the big bodies in the gravity regime to get 
involved in the cascade (which is certainly the case after $100\Myr$). As discussed above, early on the debris discs
composed of pebble piles are brighter than the discs of ``monolithic'' planetesimals.

\subsection{A steeper size distribution of large planetesimals?}
\begin{figure}
\centering
\includegraphics[width=0.49\textwidth, angle=0]{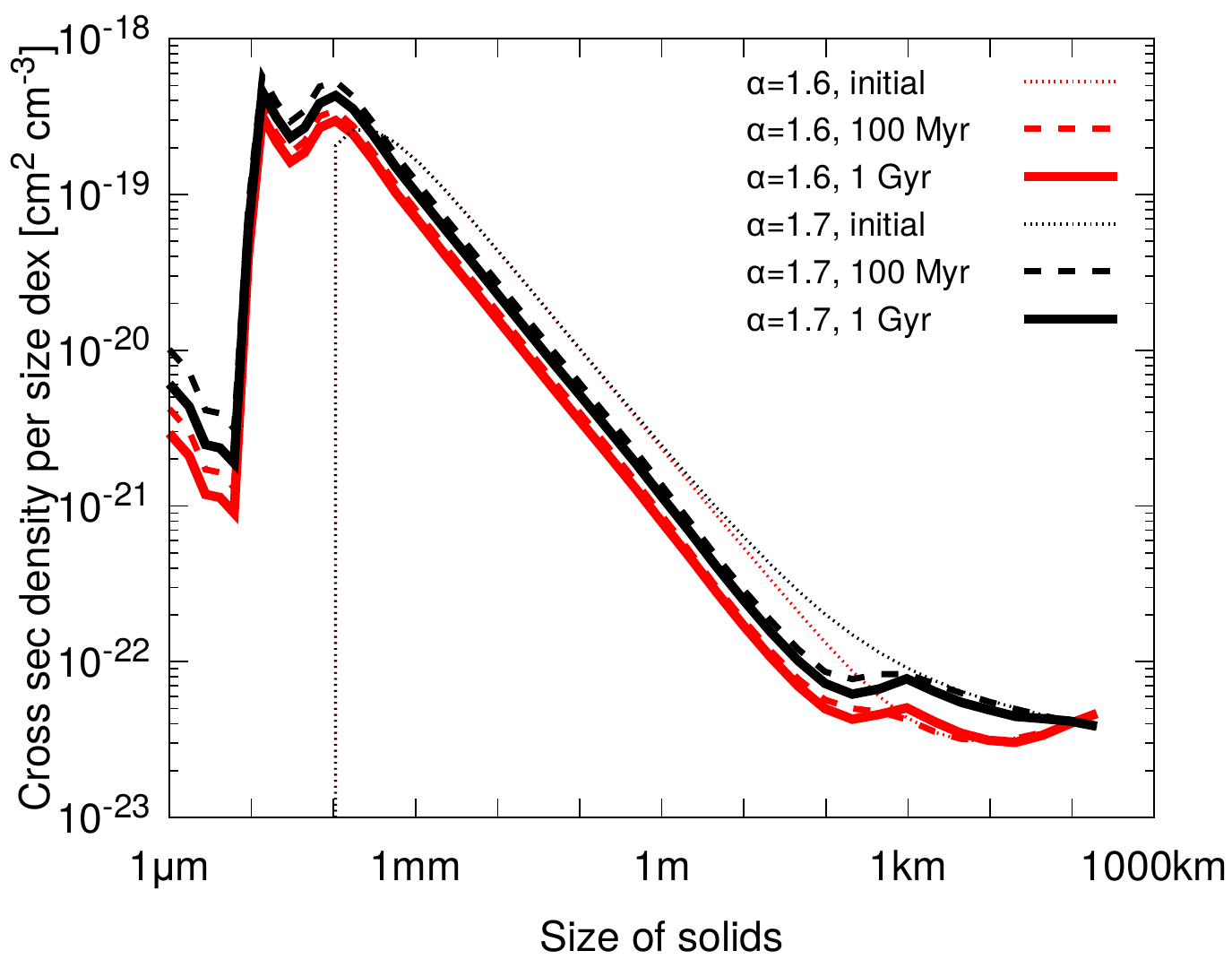}\\
\caption{
Size distributions of solids in the $M_1+M_2 = 100 M_\oplus$ and $M_1/M_2 = 1/20$ model with
a standard $Q_\text{D}^\ast$ prescription in two cases:
the reference one with $\alpha = 1.6$ (red)
and the one assuming $\alpha = 1.7$ instead (black).
Lines from thin dotted to thick solid show the distributions at $0\Myr$, $100\Myr$, and $1\Gyr$.
\label{fig:1.7}
   }
\end{figure}

The planetesimal formation models derive the slope $\alpha = 1.6$ with some uncertainty.
We checked whether a somewhat steeper slope of the population~2, namely $\alpha = 1.7$, would change
the amount of visible dust markedly. This turned out not to be the case (Fig.~\ref{fig:1.7}).
After $100\Myr$ of collisional evolution, the cross section -- and emission flux --
of the dust increases by less than a factor of two. At $1\Gyr$, it is even smaller.

\subsection{Non-collisional dust production?}
In principle, one can question a standard view that visible dust in Kuiper-belt analogs
is produced solely via collisions.
As an example, \citet{jacobson-et-al-2014} argue that for main-belt asteroids
smaller than a few kilometers in size, YORP-induced rotational disruption
significantly contributes to the erosion even exceeding the effects of
collisional fragmentation. Several erosion-like mechanisms may also enhance 
dust production.
We know that planetesimals in debris discs contain ices. For instance, the presence of $\sim 10$\% of CO ice
has been deduced from CO gas detections \citep[e.g.,][]{kral-et-al-2017c}.
The sublimation temperature of the most volatile ices such as CO, CH$_4$, and N$_2$ are lower than
$\sim 45\K$ \citep{dodson-robinson-et-al-2009b}, allowing them to sublimate at
Kuiper-belt distances from the primary stars.
Another possibility would be photodesorption of ices, which should be most efficient around A-type stars 
considered here \citep[e.g.,][]{grigorieva-et-al-2007}.
Such mechanisms of ice erosion might cause a direct, non-collisional release of dust particles from an 
icy-rocky matrix at the surfaces of planetesimals in a ``cometary'' way.

\section{Conclusions}

Our key results are as follows:
\begin{enumerate}
\item
The long-term evolution of debris discs is largely determined by the collisional strength
of solids at both ends of the size distribution, i.e., large planetesimals in the gravity regime and dust.
However, this is different at early stages of debris disc evolution, when the
system is younger than the collisional lifetime of 
planetesimals kept together by gravity.
The temporal evolution of debris discs at younger ages
is sensitive to the assumed fragmentation laws 
of intermediate-sized planetesimals, and thus to their internal structure.
Discs of ``monolithic'' planetesimals 
and those of ``pebble piles'' exhibit different evolutionary curves.
This shows a principal possibility to constrain structure and porosity of planetesimals from the observations of 
young debris discs.
However, it is not easy, since the dust fluxes also depend on other parameters such as the birth size 
distribution of planetesimals, see conclusions (ii) and (iii) below.

\item
Assuming ``monolithic'' planetesimals, as implied by the standard planetesimal formation scenario,
we are able to reproduce the observed debris disc brightness evolution. The requirement is that planetesimal
belts left after the time of the gas dispersal must include a substantial population of
sub-kilometer-sized planetesimals.
For example, a population of objects with sizes from about 1 meter to about 1 kilometer
and a Dohnanyi-like slope $\alpha \approx 1.8...1.9$ would suffice.
This population is expected in the standard planetesimal formation scenario as well,
since the growth of larger bodies in a protoplanetary disc in that scenario proceeds concurrently
with fragmentation at smaller sizes that establishes such a slope by the time of gas dispersal.

\item
For planetesimals with a pebble-pile structure, as expected in the particle concentration models,
this additional population of small planetesimals may not be required~--- and is not predicted.
Taking the initial size distribution with a slope $\alpha \approx 1.6$ between a few kilometers
and a few hundred kilometers and assuming planetesimals to be pebble piles, as suggested by these models,
we are also able to roughly reproduce the average luminosities of debris discs at all ages.

\item
While our modelling shows that the observed brightness evolution of debris discs is compatible with both
monolithic and pebble-pile planetesimals, there is a possibility to discriminate between the two cases
for the Solar System. Both the relative paucity of sub-km-sized small bodies and close-up studies of comet
67/P with Rosetta mission seem to favour the fragile, pebble-pile composition of planetesimals
in our own debris disc, supporting the ``particle concentration'' scenario of their formation.

\item
Explaining debris discs in the samples with a brightness above the average uncovers a ``disc mass problem.''
For such discs to be reproduced by collisional simulations, the total disc mass located within a $10\AU$ ring
should be on the order of $\sim 1000 M_\oplus$.
This is more than the total mass of solids
available in the protoplanetary progenitors of debris discs, which should not exceed 
a few 100s $M_\oplus$. The problem appears for all initial size spectra of 
planetesimals and their collisional strengths that we invoked in this study.

\item We consider several possibilities, one or more of which
may help to resolve the disc mass problem.
Some systems may have experienced recent major break-up of big planetesimals.
In some other systems, the collisional cascade may not have ignited immediately after the gas dispersal
(``delayed stirring'').
In still other systems, ages are quite uncertain; younger ages would imply less total mass.
It is also possible that sublimation or photodesorption of ices enhances purely collisional dust 
production 
from parent planetesimals by releasing dust from their surfaces; this would reduce the total mass needed to 
sustain the observed amount of dust as well.

\end{enumerate}

\section*{acknowledgements}
We are grateful to Mark Booth for commenting on the manuscript
and Hiroshi Kobayashi for providing us with the
details of his planetesimal growth simulations.
A speedy and insightful referee report by an anonymous reviewer is very much appreciated.
AVK, TL, and JB thank the {\it Deutsche Forschungsgemeinschaft} (DFG) for financial support through grants
Kr~2164/13-1, Lo~1715/2-1, and  Bl~298/24-1, respectively.
AJ thanks the Knut and Alice Wallenberg Foundation (grants 2012.0150, 2014.0017, 2014.0048),
the Swedish Research Council (grant 2014-5775)
and the European Research Council (ERC Consolidator Grant 724687-PLANETESYS)
for their financial support.



\newcommand{\AAp}      {Astron. Astrophys.}
\newcommand{\AApR}     {Astron. Astrophys. Rev.}
\newcommand{\AApS}    {AApS}
\newcommand{\AApSS}    {AApSS}
\newcommand{\AApT}     {Astron. Astrophys. Trans.}
\newcommand{\AdvSR}    {Adv. Space Res.}
\newcommand{\AJ}       {Astron. J.}
\newcommand{\AN}       {Astron. Nachr.}
\newcommand{\AO}       {App. Optics}
\newcommand{\ApJ}      {Astrophys. J.}
\newcommand{\ApJL}      {Astrophys. J. Lett.}
\newcommand{\ApJS}     {Astrophys. J. Suppl.}
\newcommand{\ApSS}     {Astrophys. Space Sci.}
\newcommand{\ARAA}     {Ann. Rev. Astron. Astrophys.}
\newcommand{\ARevEPS}  {Ann. Rev. Earth Planet. Sci.}
\newcommand{\BAAS}     {BAAS}
\newcommand{\CelMech}  {Celest. Mech. Dynam. Astron.}
\newcommand{\EMP}      {Earth, Moon and Planets}
\newcommand{\EPS}      {Earth, Planets and Space}
\newcommand{\GRL}      {Geophys. Res. Lett.}
\newcommand{\JGR}      {J. Geophys. Res.}
\newcommand{\JOSAA}    {J. Opt. Soc. Am. A}
\newcommand{\MemSAI}   {Mem. Societa Astronomica Italiana}
\newcommand{\MNRAS}    {MNRAS}
\newcommand{\PASJ}     {PASJ}
\newcommand{\PASP}     {PASP}
\newcommand{\PSS}      {Planet. Space Sci.}
\newcommand{\QJRAS}    {Quart. J. Roy. Astron. Soc.}
\newcommand{\RAA}      {Research in Astron. Astrophys.}
\newcommand{\SolPhys}  {Sol. Phys.}
\newcommand{\SolSysRes}{Sol. Sys. Res.}
\newcommand{\SSR}      {Space Sci. Rev.}

\input paper.bbl


\bsp    
\label{lastpage}
\end{document}